\documentclass[letterpaper,11pt]{scrartcl}
\addtokomafont{title}{}
\setkomafont{author}{\sffamily}
\setkomafont{date}{\sffamily}

\usepackage[T1]{fontenc}

\usepackage{geometry}
\usepackage[
    backend=bibtex, 
    style=chem-acs, 
    articletitle=true,
    doi=false,
    sorting=none,
    ]{biblatex}
\usepackage{graphicx}

\usepackage{amsmath}
\renewcommand*{\vec}[1]{\mathbf{#1}}

\newcommand{\azangle}{\ensuremath{\gamma}}

\newcommand*{\HtH}{\ensuremath{\rightarrow\leftarrow}}
\newcommand*{\TtT}{\ensuremath{\leftarrow\rightarrow}}

\usepackage[separate-uncertainty=true]{siunitx}

\usepackage{xcolor}

\newcommand{\beginsupplement}{
        \cleardoublepage
		\newrefsection 
        \setcounter{page}{1}
        \renewcommand{\thepage}{S\arabic{page}}%
		\setcounter{section}{0}
		\renewcommand{\thesection}{S\arabic{section}}%
		\setcounter{table}{0}
		\renewcommand{\thetable}{S\arabic{table}}%
		\setcounter{figure}{0}
		\renewcommand{\thefigure}{S\arabic{figure}}%
		\setcounter{equation}{0}
		\renewcommand{\theequation}{S\arabic{equation}}%
		}

\usepackage[noblocks]{authblk}

\newcommand{\lboro}{Department of Physics, Loughborough University, LE11 3TU Loughborough, United Kingdom}

\newcommand{\tuwien}{Institute of Applied Physics, TU Wien, Vienna 1040, Austria}
\newcommand{\ifwdresden}{Leibniz Institute for Solid State and Materials Research, Institute for Solid State Research, Helmholtzstraße 20, 01069 Dresden, Germany}
\newcommand{\tudresden}{Institute of Solid State and Materials Physics, Department of Physics, TUD Dresden University of Technology, Haeckelstraße 3, 01069 Dresden, Germany }
\newcommand{\oviedouni}{Departamento de Física, Universidad de Oviedo, 33007, Oviedo, Spain}
\newcommand{\oviedocsic}{CINN (CSIC—Universidad de Oviedo), 33940, El Entrego, Spain}
\newcommand{\soleil}{Synchrotron SOLEIL, Saint-Aubin 91190, France}
\newcommand{\uniwien}{Physik funktioneller Materialien, Universität Wien, Kolingasse 14-16, Wien 1090, Austria}
\newcommand{\alba}{ALBA Synchrotron Light Source, 08290 Cerdanyola del Valles, Barcelona, Spain}

\author[1,2]{Na\"{e}mi Leo$^{\ast,}$} 
\affil[1]{\lboro}
\affil[2]{\tuwien}

\author[3]{Daniel Wolf}
\affil[3]{\ifwdresden}

\author[4,5,6]{Alicia Estela Herguedas Alonso}
\affil[4]{\oviedouni}
\affil[5]{\oviedocsic}
\affil[6]{\alba}

\author[3,7]{Oleksandr Zaiets}
\affil[7]{\tudresden}    

\author[2]{Jakub Jurczyk}
\author[2]{Takeaki Gokita}

\author[8]{John Fullerton}
\affil[8]{Materials Science Division, Argonne National Laboratory, Lemont, IL, 60439 USA}

\author[9]{Dedalo Sanz-Hernandez}
\affil[9]{Laboratoire Albert Fert, CNRS, Thales, Palaiseau, 91767, France}

\author[10]{Claire Donnelly}
\affil[10]{Max Planck Institute for Chemical Physics of Solids, Dresden, Germany}

\author[6]{Andrea Sorrentino}
\author[6]{Eva Pereiro} 
\author[6]{Lucia Aballe} 

\author[11,12]{Peter Fischer}
\affil[11]{Materials Sciences Division, Lawrence Berkeley National Laboratory, Berkeley, CA 94720, USA} 
\affil[12]{Physics Department, University of California Santa Cruz, Santa Cruz, CA 94056, USA}

\author[13]{Rachid Belkhou}
\affil[13]{\soleil}

\author[14]{Claas Abert}
\author[14]{Dieter Suess}
\affil[14]{\uniwien}    

\author[3,7]{Axel Lubk}

\author[4,5]{Aurelio Hierro-Rodriguez}

\author[2]{Amalio Fernández-Pacheco$^{\dagger,}$} 

\newcommand*{\papertitle}{Generation of Bloch Points with Controlled Spin Texture Using Geometrical Boundary Conditions}
\title{\papertitle}

\date{$^{\ast}$n.leo@lboro.ac.uk, $^{\dagger}$amalio.fernandez-pacheco@tuwien.ac.at}

\begin{document}

\maketitle

\begin{abstract}
	Bloch points are three-dimensional topological singularities in magnetization that play a key role in topological transformations of spin textures, such as skyrmion creation or annihilation. While topology often enforces the existence of Bloch points in confined geometries like cylindrical nanowires, deterministic control over their position and magnetic configuration remains challenging. 
	Here we demonstrate the generation of Bloch points with controlled spin texture by engineering geometrical boundary conditions in three-dimensional nanomagnets. By introducing a chirality interface between two three-dimensional double-helix nanowires of opposite handedness, forming a kinked, non-collinear structure, we impose competing topological constraints that uniquely define the magnetization configuration surrounding the Bloch point. A saturating magnetic field nucleates head-to-head or tail-to-tail domain configurations at the chirality interface, producing a Bloch-point domain wall with deterministic polarity, circulation and helicity.
	This geometrical approach enables full three-dimensional control of Bloch point domain walls allowing deterministic engineering of their spin texture and its selective coupling to current-induced Oersted fields.
\end{abstract}

\section*{Keywords}
Bloch points, nano\-magnetism, electron holography tomography, magnetic soft xray tomography, magnetic chirality, helical nanowires

\section{Introduction}
Topological defects play a central role in magnetism, governing both the stability and the dynamics of nanoscale magnetic textures. Among them, Bloch points are unique as true three-dimensional singularities of the magnetization field, where the magnetization vanishes and its direction becomes undefined \cite{1965Feldtkeller,2017Feldtkeller,2014Andreas,2025Fernandez, 2025Kuchkin, 2025Yastremsky}. Bloch point singularities arise in a wide range of magnetic systems and are key to topological features of certain magnetic solitons, including chiral bobbers \cite{2018Zheng, 2022Birch}, interfaces between skyrmion tubes of different topology \cite{2019Beg,2023Lang}, and Bloch point domain walls \cite{2014DaCol, 2023Caso, 2024Tejo}. Bloch points also play a crucial role in a wide range of transient magnetization processes \cite{2019Wartelle, 2022Moreno}, such as the mobility of the aforementioned solitons \cite{2022Birch}, magnetic vortex core reversal \cite{2003Thiaville}, and the change of topological charge during the creation and annihilation of vortex-anti-vortex pairs and skyrmions \cite{2020Lib, 2022Birch}. By providing a transient singular configuration of the magnetization field, Bloch points enable topological transitions that are otherwise prohibited by the topological stability of these textures.

Although the influence of Bloch points on the field of nanomagnetism has been recognized for decades, only recent advances in nanoscale magnetic imaging have allowed to experimentally probe their 3D spin order \cite{2017Donnelly, 2020Hierro-Rodriguez} and behavior under applied fields or current pulses \cite{2025Alvaro-Gomez, 2026Gomez-Cruz}.
%
%
Regardless of this progress, the use of Bloch points for potential novel spintronic functionalities  \cite{2023Lang, 2024Sanchez, 2025Fernandez, 2025Birch, 2025Ruiz-Gomez} is still hampered by the difficulty to control their nucleation, including both their spatial position and their magnetic configuration surrounding the singularity. In particular, it remains an open question whether the spin texture around a Bloch point can be deterministically defined through designed material and geometrical engineering, rather than emerging stochastically from symmetry or defects. 

In this work, we demonstrate deterministic control of Bloch point nucleation by engineering geometrical boundary conditions, implemented via a chirality interface between 3D printed helical nanowires. Using advanced nanotomography techniques with magnetic sensitivities, including transmission electron microscopy (TEM) and magnetic soft X-ray transmission microscopy,
%
supported by micromagnetic simulations, we confirm that domain walls containing circulating Bloch points are reliably nucleated with predefined polarity, circulation and helicity angle. By imposing competing geometrical constraints at the interface, the spin texture surrounding the Bloch point is uniquely defined.
This approach, combining tailored geometry and a robust field initialization protocol, provides a pathway towards 3D spintronic device architectures containing Bloch points. 


\section{Bloch Point Singularities}
\label{sec:BPtheory}

\begin{figure}[bt]
	\centering
    \includegraphics{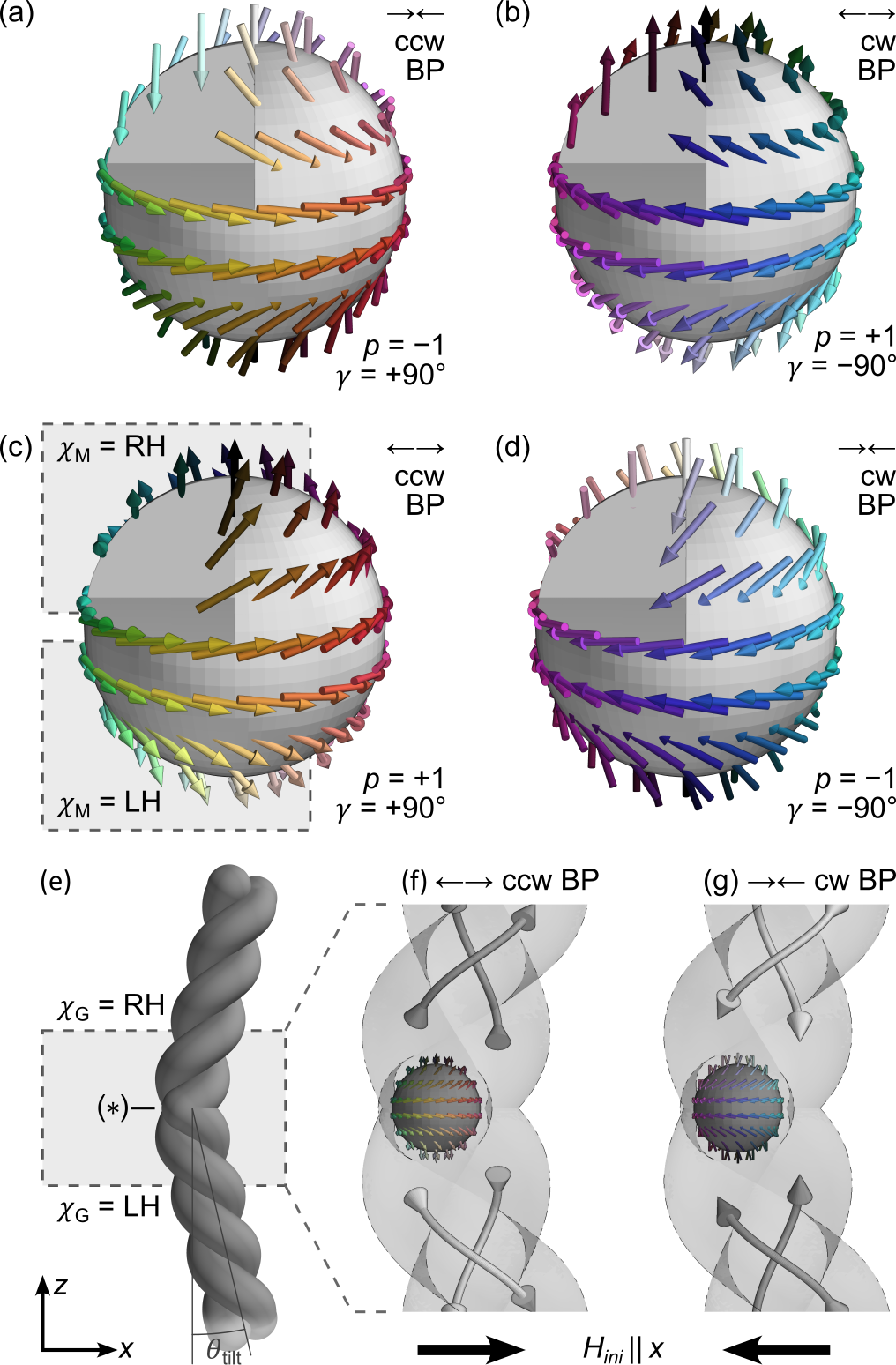}
	\caption{%
		\textbf{Geometrical control of Bloch point spin texture using a chirality interface.} 
		(a-d)~Circulating Bloch points with varying polarity $p$ (head-to-head \HtH, $p=-1$ and tail-to-tail \TtT, $p=+1$) and helicity $\azangle$ describing clockwise (cw, $\gamma=-\ang{90}$) and counter-clockwise (ccw, $\gamma=+\ang{90}$) spin circulation. These parameters define the spin state around the Bloch point singularity. 
		While each half of the spin texture around a Bloch point has a well-defined magnetic chirality $\chi_M$, as shown in (c), the net Bloch point spin texture is achiral. 
		(e)~3D nanowire connecting a left-handed (LH, bottom) with a right-handed (RH, top) double helix, with each segment tilted by $\pm\theta_\text{tilt}$ away from the $z$ direction. The chirality interface is marked with ($\ast$) and imposes competing geometrical boundary conditions.
		(f,g)~Field-induced initialization of Bloch points after saturation with a sufficiently strong field $\mathbf{H}_\text{ini}\parallel x$. The sign of $H_\text{ini}$ sets the Bloch point polarity, while its circulation is determined by the geometrical chirality $\chi_G$ on either side of ($\ast$), resulting in Bloch points with controlled spin texture.
        %
		}
	\label{fig:BPs}
\end{figure}

%



The micromagnetic state in the vicinity of an axially symmetric Bloch point can be described by an analytical model \cite{2012Pylypovskyi,2024Hermosa-Munoz} using spherical coordinates, expressing the local direction of the reduced moment $\vec{m}(\vec{r}) =(m=1, \theta_m, \phi_m)$ at position 
$\vec{r}=(r, \theta, \phi)$:
\begin{eqnarray}
	\theta_m(\theta) & = & p\theta + \frac{\pi}{2} \left(1 - p\right) \, ,\label{eq:theta_m} \\ 
	\phi_m(\phi) & = & v\phi + \azangle \, . \label{eq:phi_m}
\end{eqnarray}
Here, the discrete parameters $p$, $v$, and the continuous angle \azangle\ are key quantities that determine the static and dynamic properties of the Bloch point \cite{2012Pylypovskyi}.
%
The discrete polarity $p$ determines whether a tail-to-tail (\TtT, $p=+1$) or head-to-head (\HtH, $p=-1$) configuration is observed when crossing the equatorial plane. The discrete vorticity $v$ determines whether the spin state within the equatorial plane is a vortex ($v=+1$) or an antivortex state ($v=-1$). Finally, the continuous helicity angle $\gamma$ describes how moments at the equator point either inward ($\gamma=\ang{0}$) or outward ($\gamma=\ang{180}$) for a hedge-hog-like Bloch point, or rotate around the central axis ($\ang{0} < |\gamma |< \ang{180}$) in a twisted Bloch point configuration. 

A subset of twisted Bloch points are the four vortex-like \textit{circulating} Bloch points depicted in Fig.~\ref{fig:BPs}(a-d), for which $v=+1$, $p=\pm1$ and $\gamma=\pm\ang{90}$.
%
When viewed from above, these Bloch points exhibit a well-defined counter-clockwise [ccw, $\gamma=\ang{+90}$, Fig.~\ref{fig:BPs}(a,c)] or clock-wise [cw, $\gamma=\ang{-90}$, Fig.~\ref{fig:BPs}(b,d)] circulation in their equatorial plane.
This circulation sense is degenerate in highly symmetric systems. For instance, in cylindrical nanowires clockwise and counter-clockwise Bloch points appear with equal probability \cite{2019Wartelle,2022Moreno,2023Hermosa}.
%
%
%
A uniformly chiral geometry (e.g., a helix with fixed handedness) likewise does not uniquely select the circulation of a Bloch point, since both clockwise and counter-clockwise configurations remain equivalent. 
%
This reflects the fact that a Bloch point connects regions of opposite magnetic chirality:
As shown in Fig.~\ref{fig:BPs}(c), the magnetic chirality $\chi_M$ -- which can be understood as the product between the sense of circulation and the sign of the axial magnetization (also called vortex polarization in Ref.~\cite{2025Fullerton}) -- of the top ($\chi_M$ right-handed, RH) and bottom half ($\chi_M$ left-handed, LH) of the spin state cancel each other out, i.e., $\chi_M^\text{top}=-\chi_M^\text{bottom}$, i.e., the net Bloch point spin texture is achiral.
Deterministic selection is only achieved when chirality is imposed \textit{locally}, for example at an interface that breaks the chirality equivalence.



Tailored helical nanostructures grown by recently-established 3D nanoprinting techniques \cite{2017Fernandez-Pacheco,2017Pablo-Navarro,2017SanzHernandez,2020Skoric} enable interfacing opposing geometric chirality regions for the controlled nucleation of circulating Bloch points. 
As shown in previous works \cite{2020Sanz-Hernandez,2025Fullerton}, such double-helix nanowires allow to control the magnetic chirality through their geometric chirality, i.e., $\chi_M=\chi_G$, thereby enabling the controlled creation of metastable spin structures such as chiral domain walls \cite{2020Sanz-Hernandez} and skryrmion tubes \cite{2025Fullerton} by simple magnetic-field protocols.

The proposed structure for controlled Bloch point nucleation is shown in Fig.~\ref{fig:BPs}(e), and consists of double helices of opposite geometric chirality $\chi_G$, here a left-handed segment at the bottom and a right-handed segment at the top, meeting at the chirality interface marked with ($\ast$). 
Each segment is tilted by an angle of $\pm\theta_\text{tilt}$ away from the longitudinal $z$ axis. This tilt towards the $x$ axis allows initialization of a head-to-head or tail-to-tail domain wall at the chirality interface with a sufficiently strong transverse magnetic field $\mathbf{H}_\text{ini}\parallel x$ \cite{2023Hermosa}. 
Subsequent relaxation of the spin structure, together with the imprinted opposing geometric chirality on both sides of the interface, supports the stabilization of a central circulating Bloch point at remanence. 

Importantly, there is a direct coupling between the magnetostatic charge and the circulation of the created Bloch point, determined by the change of geometric chirality at $(\ast)$ and by the sign of $H_\text{ini}$ in this system. In our case, interfacing left- and right-handed double helices stabilizes either a tail-to-tail counter-clockwise or a head-to-head clockwise Bloch point domain wall, as shown in Figs.~\ref{fig:BPs}(f,g). 
%
Therefore, this geometrical interface-induced mechanism enables simultaneous control over both the Bloch point polarity (head-to-head or tail-to-tail, set by the direction of the initially applied field), its circulation (set by the RH/LH or LH/RH order of the helical segments) and its position (localized close to the chirality interface). This level of control is not achievable in cylindrical nanowires, where the position of Bloch points can be influenced by curvature \cite{2025Ruiz-Gomez}, yet both tunability of its spin texture and spatial determinism 
remains limited \cite{2019Wartelle,2022Moreno,2023Hermosa}. The key enabling factor here is the connectivity of two double-helix nanowires via the chirality interface (*), a topological structural defect which imprints opposite geometric chirality $\chi_G$ across the interface, and thus to a corresponding opposite magnetic handedness $\chi_M$.

\section{Results}
\label{sec:results}

\begin{figure}[bt]
	\centering
	\includegraphics{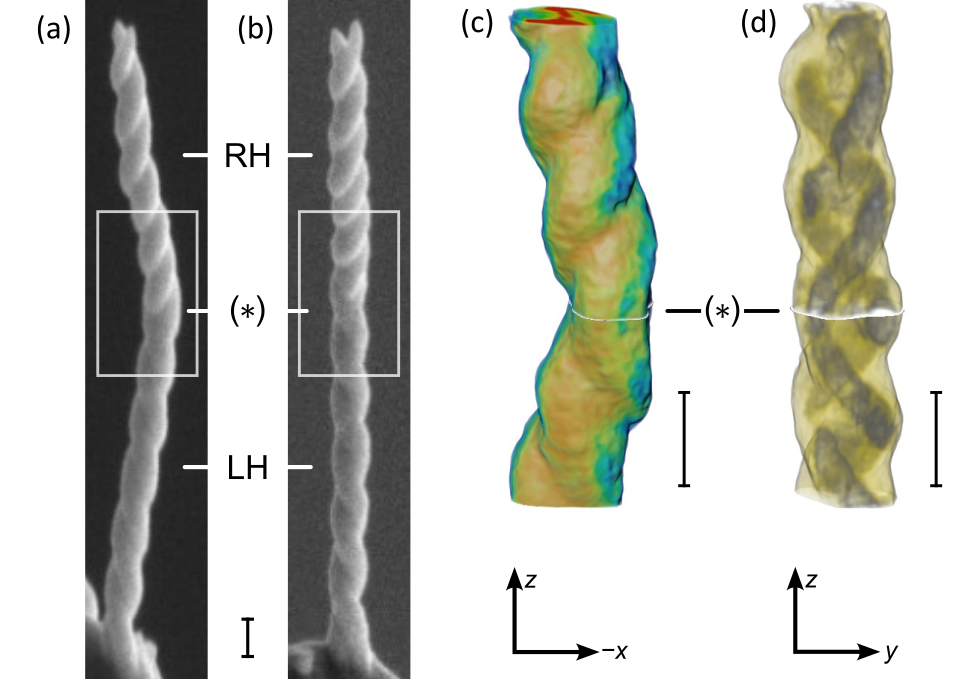}
	\caption{%
		\textbf{3D nanostructure defining the geometrical boundary conditions for Bloch point generation.} 
		(a)~Side and (b)~front view of FEBID-grown Co double-helix structures, combining a bottom LH and top RH section tilted by $\pm\ang{15}$. 
		(c)~Opaque and (d)~transparent volume rendering of the nanowire's mean inner potential (proportional to material contrast) reconstructed by electron holographic tomography. 
		The position of the chirality interface is marked by $(\ast)$ and defines the boundary where Bloch points are nucleated and their topology is imposed.
		Scale bars measure \SI{100}{nm}.
		}
	\label{fig:structure}
\end{figure}

\subsection{Tailoring the Geometric Chirality Interface}

We used focused-electron beam-induced deposition (FEBID) \cite{2020Skoric} to 3D nano-print free-standing magnetic Co helical nanowires \cite{2020Sanz-Hernandez, 2025Fullerton} (Methods Sec.~\ref{sec:methods:FEBID}). Scanning electron microscopy (SEM) images in Fig.~\ref{fig:structure}(a,b) show the resulting structures in side and front view. 
%
%
In the middle of the structure, marked with $(\ast)$, the geometric chirality of the double helix sections changes from $\chi_G^\text{bottom}=\text{LH}$ to $\chi_G^\text{top}=\text{RH}$. 
Each chiral section is inclined by $\theta_\text{tilt}=\pm\ang{15}$, with the value of the tilt angle being a compromise of three considerations:
First, \ang{15} is sufficiently large for a transverse field $H_x$ to initialize the head-to-head or tail-to-tail configuration required to stabilize a circulating Bloch point. 
Second, considering FEBID growth, larger tilt angles result in overhangs that are difficult to realize, and furthermore promotes electron scattering that can lead to undesirable co-depositions. 
Third, as indicated by micromagnetic simulations, a tilt angle of \ang{15} is small enough to suppress flux leakage at the chirality interface, which would favor the formation of a transverse domain wall instead of a Bloch point domain wall.


We performed TEM-based off-axis electron holographic tomography, a combination of electron holography and electron tomography to obtain a three-dimensional reconstruction of the electrostatic potential \cite{2015Wolf, 2019Wolf} (Methods Sec.~\ref{sec:methods:TEM}). 
The electrostatic potential arises from the averaged atomic potentials of the material over a volume of a few cubic nanometers (for crystalline materials referred to as mean inner potential), and thus can be used to visualize the material contrast and the 3D morphology of the nanostructure.

Fig.~\ref{fig:structure}(c) shows the 3D morphology of the double-helix structure in the vicinity of the chirality interface, rendered with a voxel size of (\SI{0.8}{nm})$^3$.
The nanostructure shape revealed closely resembles the targeted 3D nano-printed CAD geometry shown in Fig.~\ref{fig:BPs}(e). The chirality interface at $(\ast)$, where the bottom LH helix merges into the top RH helix, can be easily identified and is marked with a white outline.
%
%
Figure~\ref{fig:structure}(d) depicts a transparent 3D volume highlighting the inner structure of the nanowire in gray to emphasize the interwoven strands deposited during the layer-by-layer 3D printing process. A 3D animation and further information are provided in the Supporting Information.

%
%
%

A reduced diffraction contrast observed in bright-field TEM images (not shown) 
suggests that individual Co grains are typically smaller than \SI{10}{nm}, as previously reported for this FEBID material \cite{2025Fullerton}. The obtained nanocrystalline structure indicates that the magnetic anisotropy primarily arises from the imprinted nanowire geometry.
This shape anisotropy is crucial because it establishes the strong coupling between magnetic and geometric chirality, i.e.,  $\chi_M=\chi_G$. 

\begin{figure*}
	\centering
	\includegraphics{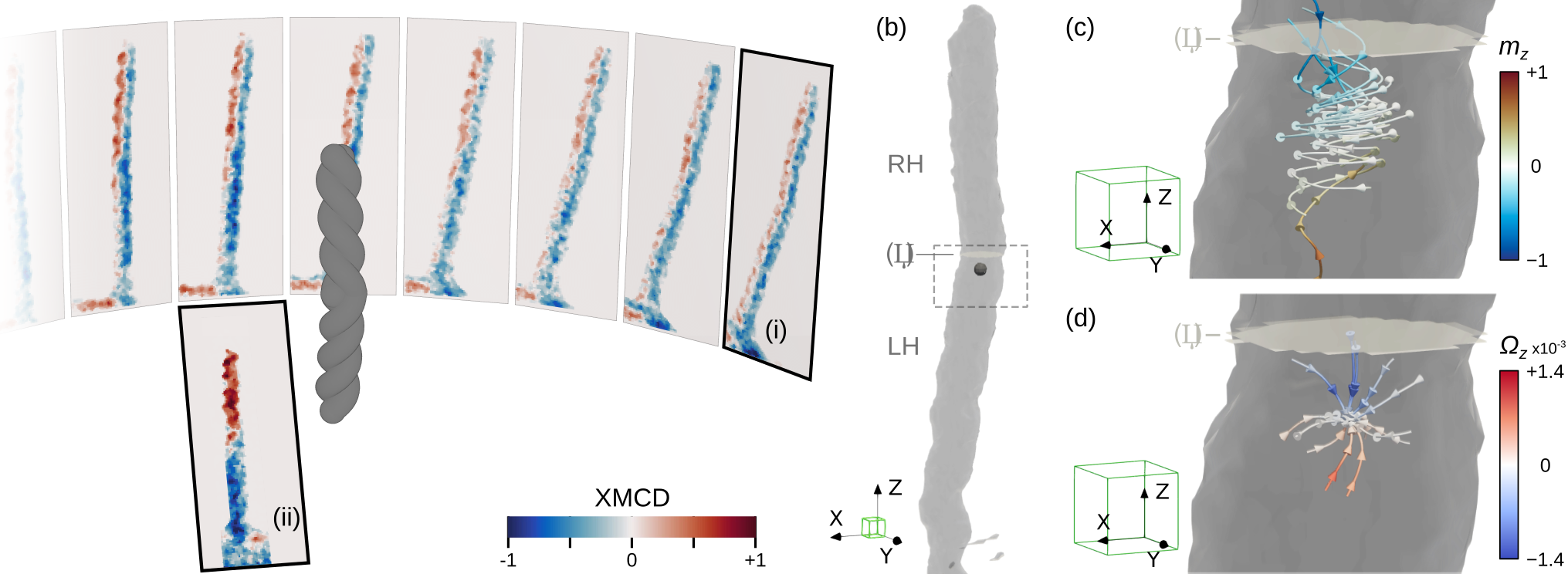}
	\caption{%
		\textbf{Direct observation of Bloch points with controlled spin texture by XMCD tomography.}
		(a)~Selection of XMCD images taken under different rotation and tilt angles for the reconstruction of the 3D magnetization. 
        The XMCD contrast in the highlighted panels indicates (i)~uniform circulation across the length of the nanostructure, and (ii)~opposite axial magnetization in both helical segments, implying the presence of a Bloch point.
		%
		(b)~Reconstructed electronic structure with voxel size of (\SI{10}{nm})$^3$. The chirality interface plane is highlighted with ($\ast$). The region of interest around the Bloch point (gray sphere) shown in (c,d) is highlighted (dashed line).
		(c)~Reconstructed magnetization $\vec{m}_\mathrm{sol}(\vec{r})$ of a clockwise-circulating head-to-head Bloch point, directly revealing a Bloch point with well-defined polarity and circulation. 
		%
		(d)~Magnetic vorticity field $\vec{\Omega}(\vec{r})$, showing that the Bloch point acts as a sink for the emergent field and confirming its topological character. 
		%
		Scale cubes measure (\SI{50}{nm})$^3$. Streamlines and arrows denote the local direction of the vector fields $\vec{m}(\vec{r})$ and $\vec{\Omega}(\vec{r})$, and the color the magnitude of the axial components $m_z$ and $\Omega_z$, respectively.
        %
		}
	\label{fig:xmcd-tomo}
\end{figure*}

\subsection{Direct Observation of Bloch Points by 3D Magnetic Tomography}


To directly demonstrate the existence of Bloch points in our system, we employ magnetic soft x-ray nanotomogaphy at the MISTRAL beamline of the ALBA synchrotron \cite{2015MISTRAL}. Using the x-ray magnetic circular dichroism (XMCD) signal from the at Co $L_2$ absorption edge, which measures the projection of the magnetization onto the photon propagation direction, and different sample orientations, we can obtain a volume-resolved reconstruction of the vector character of the local magnetic moments.
As a result, the measurements provide direct access to the three-dimensional spin texture surrounding the Bloch point down to the spatial resolution limit of about \SI{50}{nm}. 

The Bloch point state was initialized by applying a transverse field of about \SI{1}{T} along the kink of the nanowire, following the scheme described in Fig.~\ref{fig:BPs}(f-g) (see Methods Sec.~\ref{sec:methods:alba}).
To perform a vector magnetic tomographic reconstruction \cite{2017Donnelly,2018HierroRodriguez}, the results of two measurement series were combined -- one, were the double-helix nanostructure is rotated around its main axis from \ang{-65} to \ang{+64} in \ang{1.5} steps, as well as getting tilted towards the beam from \ang{-33} to \ang{+19} in \ang{1} steps (see Methods Sec.~\ref{sec:methods:alba} and Supplemental Material Sec.~\ref{ssec:methods:alba} for further details). 

Figure~\ref{fig:xmcd-tomo}(a) shows representative XMCD measurements for the rotation (horizontal images) and tilt (vertical images) series.
%
%
From the highlighted XMCD images in Fig.~\ref{fig:xmcd-tomo}(a) one can deduce: (i)~a magnetic state of uniform circulation throughout the nanostructure; 
and (ii)~opposite axial magnetization at the two sides of the chirality interface, \textit{i.e.,} for the two segments of opposite geometric chirality. 
%
Figure~\ref{fig:xmcd-tomo}(b) shows the reconstructed structural contrast 
of the double helix geometry with the central kink corresponding to the geometrical chirality interface highlighted by a white plane. 
Analyzing the magnetization reconstruction in the vicinity of the chirality interface (Fig.~\ref{fig:xmcd-tomo}(c)), we confirm the presence a head-to-head Bloch point with clockwise circulation (when viewed from above), in agreement with our expectations from Fig.~\ref{fig:BPs}(g). 

Interestingly, the singularity position, marked with a black sphere in Fig.~\ref{fig:xmcd-tomo}(b), is offset by about $\SI{60}{nm}$ from the chirality interface [$(\ast)$, white plane] towards the bottom left-handed segment. 
Recent results have shown that Bloch point domain walls are energetically unfavorable in sections of high geometric curvature \cite{2025Ruiz-Gomez}. Here, the kink introduces a localized curvature maximum at the chirality interface, therefore is expected to play an important role in the energetic landscape and pinning of the spin texture around the Bloch-point.

Considering the spin dynamics governed by the damping term in the Landau-Lifschitz-Gilbert (LLG) equation, the displacement of a Bloch point domain wall from a region of high curvature is predicted to depend on the Bloch point polarity (see Supporting Information Sec.~\ref{ssec:curvature_driven_displacement}). Specifically, analytical arguments and micromagnetic simulations show that upon initialization from a saturated state, a head-to-head (tail-to-tail) Bloch point moves towards the left-handed (right-handed) chiral section.
%
The displacement observed in the XCMD data agrees with this prediction. However, as shown below by the higher-resolution TEM measurements, the Bloch point position is also strongly influenced by local pinning sites and microstructural inhomogeneities, leading to deviations from this idealized behavior. 
%
The displacement of the Bloch point from the chirality interface at $(\ast)$ also has important consequences for the local magnetic configuration, as it creates a small segment in which the condition $\chi_M=\chi_G$ is not fully obeyed. As we discovered in a previous work \cite{2025Fullerton}, in this section a metastable fractional skyrmion-like tube state 
is stabilized. In the present case, this spin texture terminates at the Bloch point.

In Fig.~\ref{fig:xmcd-tomo}(d) we represent streamlines of the magnetic vorticity $\vec{\Omega}(\vec{r})$ \cite{2021Donnelly} calculated from the reconstructed magnetization [Eq.~(\ref{eq:methods:ALBA:emergent_field}) in Methods]. The vorticity $\vec{\Omega}(\vec{r})$ has the same functional form as the emergent (topological) magnetic field and differs only by a constant prefactor.
%
%
%
For the head-to-head magnetic configuration observed here, the Bloch point acts as a sink for the emergent field, as indicated by the arrows on the streamlines of $\vec{\Omega}(\vec{r})$ in Fig.~\ref{fig:xmcd-tomo}(d), whereas a tail-to-tail Bloch point would act as a source. This source–sink behavior reflects the effective monopole-like character of the Bloch point in the emergent field and its associated non-trivial topology.

Finally, the XMCD vector reconstruction allows us to estimate the helicity angle $\azangle$ in Eq.~(\ref{eq:phi_m}), providing access to the third parameter describing the spin texture around a Bloch point. From the experimental data, we found $\azangle\approx\ang{100}$, in good agreement with simulations (Supporting Information Sec.~\ref{ssec:helicity_angle}). This value is set by the local geometry of the system, in this case the kinked double-helix with a chirality interface. The value $\azangle>\ang{90}$ means that spins in the equatorial plane of the Bloch point not only circulate but also twist inwards, creating a Bloch point of slightly hyperbolic character \cite{2024Hermosa-Munoz}. 
%
%

\subsection{High-Resolution Magnetic Imaging by Electron Holography}

\begin{figure*}[bt]
	\centering
	\includegraphics[width=\textwidth]{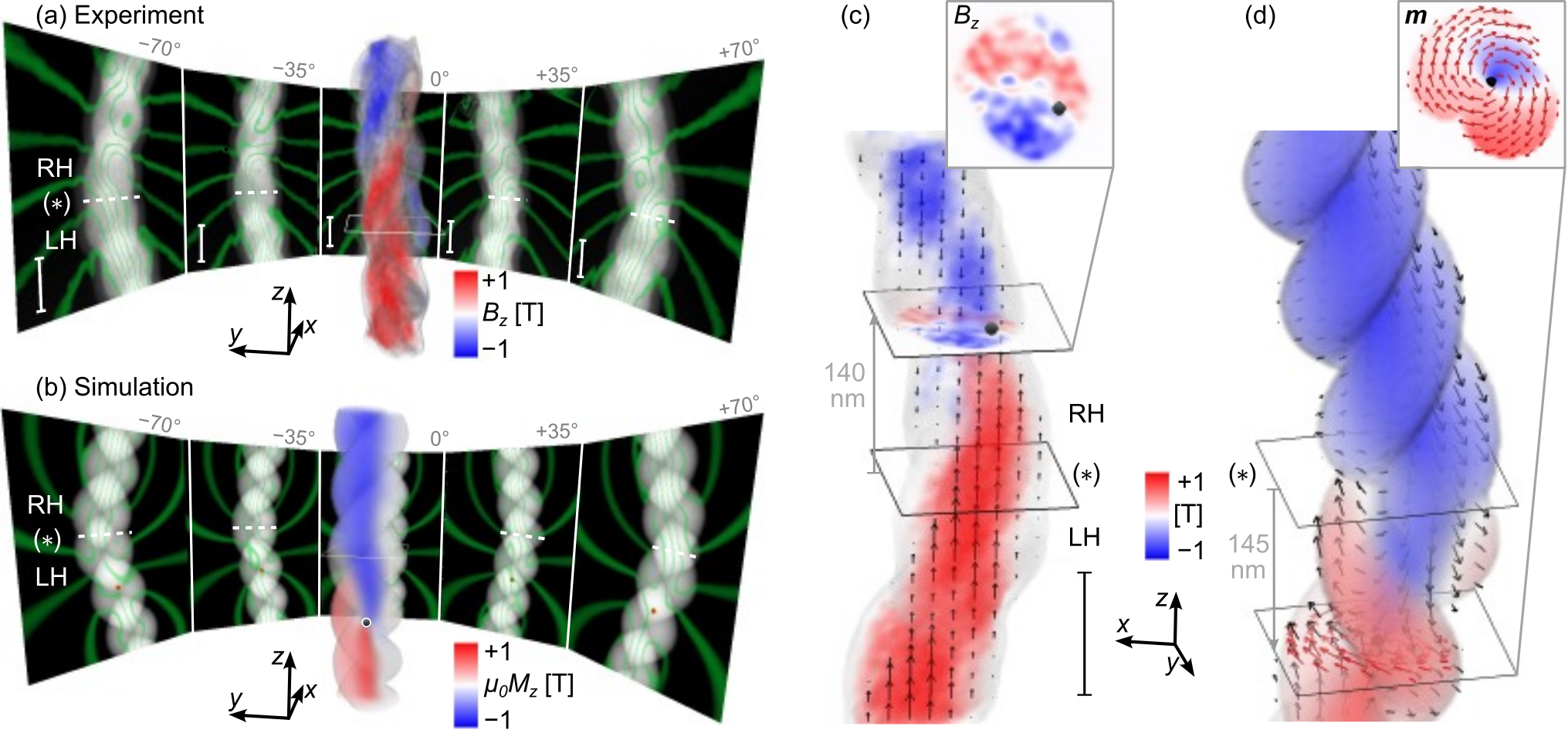}
	\caption{
		\textbf{Stray-field and internal magnetic induction revealing details of Bloch point domain walls.}
		(a)~Experimental and (b)~simulated phase images after initialization of a Bloch-point state, viewed under different projection angles. The electric and magnetic contributions are shown in grayscale and as green iso-lines, respectively. The chirality interface $(\ast)$ is marked by dashed lines and cut planes. The monopole-like symmetry of the stray-field patterns is consistent with the presence of a Bloch point.
		(c)~Tomographic reconstruction of the internal magnetic induction component $B_z(\vec{r})$ [also shown in the foreground of (a)], showing the domain wall associated with the Bloch point. The inferred position of the singularity is marked with a black dot (see inset plane).
  	(d)~Magnetization state $\vec{m}(\vec{r})$ obtained from micromagnetic simulations, with the inset showing the circulation in the equatorial plane of the Bloch point (black dot), confirming the circulating texture of the Bloch point.
		Scale bars measure \SI{100}{nm}.
		%
		%
	}
	\label{fig:Fig2-tem-tomo}
\end{figure*}

Having established the presence of Bloch points through direct XMCD tomography, we now employ electron holography and electron holographic tomography to obtain higher-resolution insight into the magnetic structure at the chirality interface. To this end, we investigate a similar nanostructure as before, initializing the Bloch point with a multi-step field protocol within the transmission electron microscope; while the experimental protocol follows the same principle as the initialization shown in Fig.~\ref{fig:BPs}(e,f) and used in the XMCD experiments (Methods Sec.~\ref{sec:metods:TEM:init}). Electron holographic tomography reconstructs, in addition to the electrostatic potential, the three-dimensional distribution of components of the magnetic induction $\vec{B}$ within \cite{2015Wolf, 2019Wolf} and outside \cite{2014Lubk} a 3D nanostructure at nanometer resolution. 
In a single-axis tomography experiment, we reconstruct the component of the magnetic induction parallel to the rotation axis only. 
%


Fig.~\ref{fig:Fig2-tem-tomo}(a) shows the combined experimental dataset of the external and internal magnetic state at multiple angles measured at remanence.
These results are compared to micromagnetic simulations shown in Fig.~\ref{fig:Fig2-tem-tomo}(b) (Methods Sec.~\ref{sec:methods:sims}).
Projections taken from different directions show electrostatic phase maps in gray superimposed with magnetic equi-phase lines in green representing magnetic field lines transversal to the electron beam direction. A volume rendering in the center shows the reconstructed magnetic induction $B_z(\vec{r})$ within the nanostructure, indicating the head-to-head domain configuration with a blue-red contrast. 

Both experimentally measured as well as simulated stray field lines show a high degree of symmetry, with little change under different viewing angles. A directional variation of the far field would indicate leakage of magnetic flux, as is, for example, observed for transverse domain walls \cite{2013Biziere} (see also Supplemental Materials Sec.~\ref{ssec:methods:tem} and Fig.~\ref{sfig:transverse-wall-TEM-contrast} for a comparison). 
The lack of variation in the field line pattern with viewing direction therefore indicates a monopole-like far-field stray-field distribution, consistent with the presence of a Bloch point domain wall. 
Furthermore, for the simulated patterns shown in Fig.~\ref{fig:Fig2-tem-tomo}(b) the field lines originate from the position of the Bloch point (indicated with a black sphere). 
%
%
The agreement between the measured and simulated datasets, together with the symmetry of the far-field lines which indicate an underlying monopolar character, strongly supports the presence of a Bloch point in the experimental data, in full consistency with the XMCD tomography results discussed above.

%
%

Figs.~\ref{fig:Fig2-tem-tomo}(c,d) compare the volume rendering of the internal induction $B_z(\vec{r})$ from the experimental reconstruction to the 3D spin state $\mu_0\vec{m}(\vec{r})$ from micromagnetic simulations in greater detail.
%
Red and blue indicate volumes where experimental $B_z$ and simulated $m_z$ point upward and downward, respectively, with a head-to-head domain wall in-between. The boundary between $\pm B_z$ volumes extends about \SI{250}{nm} (i.e., about one helix periodicity), and is inclined in the nanowire (see also Supporting Information Fig.~\ref{sfig:TEM_Tomo_regions}.

Micromagnetic simulations in Fig.~\ref{fig:Fig2-tem-tomo}(d) show comparable results to the experiment, as well as demonstrate a well-defined circulation across the length of the nanostructure (see insets). 
%
Furthermore, features in magnetic phase images computed from $m_z(\vec{r})$ simulations under certain angles, such as the anti-vortex like textures shown in Fig.~\ref{fig:Fig2-tem-tomo}(b), can be inferred to projections of the magnetic induction in the vicinity of the Bloch point. Since these features are also observed in experimental phase images, they can be used to triangulate the 3D position of the Bloch point from at least two different projection angles (Supporting Information Sec.~\ref{ssec:methods:tem}). The Bloch point position is indicated with a black dot in the experimental data shown in Figs.~\ref{fig:Fig2-tem-tomo}(c). Along the axial $z$ direction, the deduced Bloch point position also coincides with the $xy$ plane where the average axial magnetization $\left<B_z\right>_{xy}$ is minimal (see inset in Figs.~\ref{fig:Fig2-tem-tomo}(c,d)). Inspecting the latter, we attribute the differences between the experimental and simulated cross‑sections to the fact that the position of the Bloch point is more off-center in the experiment than in the simulation.

\begin{figure}[tb]
	\centering
	\includegraphics{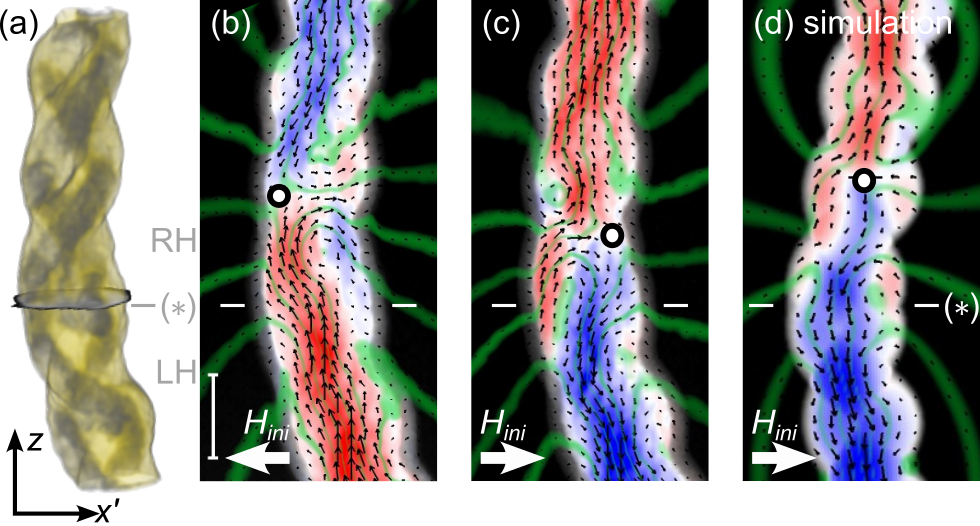}
	\caption{%
		\textbf{Deterministic control of Bloch point polarity and position.}
		(a)~Volume rendering of the structural contrast around the chirality interface $(\ast)$.
		(b,c)~Maps of external (green lines) and internal magnetic induction $\left<B_{x^\prime z}(x^\prime, z)\right>_{y^\prime}$ (arrows, blue/red denotes the magnitude of the $B_z$ component) at remanence, after initialization with opposite $H_\text{ini}$, demonstrating controlled switching between head-to-head and tail-to-tail Bloch point configurations.
		%
		(d)~Equivalent induction map calculated from micromagnetic simulations, showing the state equivalent to (c).
		Inferred positions of the Bloch point are marked with black circles; these always shows a displacement away from the chirality interface~$(\ast)$ due to curvature and pinning effects.
		Scale bar in (b) measures 100~nm.
        %
		%
	}
	\label{fig:Fig2-TEM-yz}
\end{figure}

The spin texture surrounding the Bloch point can be changed by the field initialization protocol. 
This is demonstrated in Figure~\ref{fig:Fig2-TEM-yz}, with Fig.~\ref{fig:Fig2-TEM-yz}(a) showing the visualized region around the chirality interface. Figs.~\ref{fig:Fig2-TEM-yz}(b-d) represent the projected magnetic induction maps as arrow plots overlayed with green field lines, and the axial $B_z$ component of the nanowire shown in red-blue.
%
Figures~\ref{fig:Fig2-TEM-yz}(b) and (c) compare the magnetic state obtained after initialization with opposite $H_\text{ini}$. The change in red/blue contrast clearly demonstrates the formation of a head-to-head and tail-to-tail state, respectively. The circulation is not directly resolved due to the averaging over depth, but the good agreement with micromagnetic simulations, shown in Fig.~\ref{fig:Fig2-TEM-yz}(d), allows us to identify clockwise head-to-head and counter-clockwise tail-to-tail Bloch point configurations, as predicted in Figs.~\ref{fig:BPs}(f,g).

In Figs.~\ref{fig:Fig2-TEM-yz}(b-c) the Bloch point positions are marked with black circles, and in all cases the Bloch point is displaced along $z$ away from the chirality interface $(\ast)$ [dashed line] by \SI{70}{nm} to \SI{140}{nm}.
%
Although there is overall good agreement with the micromagnetic simulations, we find a discrepancy in that the Bloch point is found to be consistently pinned on the RH segment for all TEM measurements. This finding is in contrast to the above discussed predicted behavior for the Bloch point to displace according to its polarity.
%
We ascribe this discrepancy to pinning 
caused by the  variation in density of magnetic Co (as indicated in Fig.~\ref{fig:structure}(d), and in contrast with the homogeneous material assumed in the micromagnetic simulations), as well as possible field misalignments during the multi-step Bloch point initialization procedure in the TEM experiments (Methods Sec.~\ref{sec:metods:TEM:init}).



In addition to an axial displacement away from the chirality interface, the Bloch point is also found to shift within the perpendicular $(x,y)$ plane towards the outer surface of the structure. 
%
This transverse displacement can be related to competing micromagnetic energy contributions. While magnetostatic energy favors configurations that minimize stray fields and thus a more centered position, exchange energy penalizes the strong magnetization gradients associated with the Bloch point. In the curved geometry of the system, these gradients can be partially relieved by shifting the Bloch point toward the outer surface, leading to an off-center equilibrium position. Cobalt density variations
[Fig.~\ref{fig:structure}(d)] may further favor the displacement of the Bloch point toward the surface.
The superior spatial resolution of TEM holography therefore allows to discern effects of the local structural environment on the location and behavior of a Bloch point.

The combined use of x-ray magnetic tomography, electron holographic tomography, and micromagnetic simulations allows us to probe the Bloch-point spin texture across different length scales, yielding a consistent three-dimensional picture of the magnetization around the Bloch point.


\section{Conclusions}
\label{sec:conclusions}

Bloch points -- three-dimensional singularities where the magnetization vanishes -- represent unique magnetic topological defects \cite{1965Feldtkeller,2017Feldtkeller}. 
%
However, strategies for their controlled nucleation have only recently emerged, e.g., using curvature in cylindrical nanowires as a geometrical mechanism to nucleate Bloch points with well-defined polarity \cite{2023Hermosa}. 

Here, we extend the paradigm of 3D curvature for deterministic Bloch point nucleation in chiral nanowires with double-helix geometry, where the intrinsic chirality of the helical wires provides direct control over key internal degrees of freedom of circulating Bloch points, namely their circulation, helicity and polarity. 
By interfacing left- and right-handed helices, we introduce a chirality interface where the geometric chirality reverses, which in turn acts as a robust nucleation site for Bloch point domain walls. The deliberate curvature introduced at the chirality interface further enables initialization of either head-to-head or tail-to-tail Bloch point walls under applied magnetic fields.
Careful tailoring of the nanostructure thus establishes a direct correspondence between geometrical topology (i.e., reversal of geometric chirality) and magnetic topology (i.e., emergence of Bloch point singularities). Importantly, the chirality of the nanowires imprints a well-defined circulation on the Bloch points, in contrast to high-symmetry cylindrical nanowires or nanospheres \cite{2012Pylypovskyi,2019Wartelle,2022Moreno,2023Hermosa}, where left-handed and right-handed circulations are energetically degenerate. 

%
Our approach is conceptually analogous to previous theoretical proposals of interfacing films with opposite sign of the Dzyaloshinskii-Moriya interaction (DMI) to stabilize circulating Bloch points \cite{2019Beg, 2022Lang, 2025Winkler}. However, engineering such finely balanced thin-film layers can be challenging, as both sign and magnitude of the DMI will need to be tuned with high accuracy. Furthermore, thin-film systems currently lack practicable protocols to initialize and address Bloch points in a reliable manner, and instead rely on intrinsic defects as nucleation sites \cite{2018Donnelly, 2021Donnelly_a, 2024Hermosa-Munoz}.

In contrast, by experimentally controlling the geometric -- and thus magnetic -- chirality in a tailored 3D geometry made from a single material, we here demonstrate full control over the spin texture of circulating Bloch points.
In addition to the Bloch point initialization using high magnetic fields used in this work, we also identify an alternative field protocol using lower field magnitudes that is suitable for in-situ studies (Supporting Information Sec.~\ref{ssec:alternative_init_protocol_SOLEIL}). 
%

This geometrical approach to nucleate Bloch points with tailored spin textures provides a powerful platform for experimental exploration of Bloch point physics, including their high-frequency response \cite{2019Im, 2024Tejo, 2025Winkler} and non-reciprocal interactions with current-generated Oersted fields \cite{2019Wartelle, 2023Lang, 2025Birch, 2026Gomez-Cruz}. 
Importantly, the deterministic relation between Bloch-point polarity and circulation achieved here provides direct control over the coupling of the domain wall to the current-induced Oersted field, enabling selective actuation depending on the internal state. Geometric selection and control of Bloch points therefore holds strong potential for future Bloch-point based spintronic devices, including racetrack memory and logic architectures \cite{2022Lang, 2024Sanchez, 2025Fernandez}.


\section{Methods}

\subsection{Sample Growth}
\label{sec:methods:FEBID}

Magnetic Co double helix structures were grown on pre-cut Cu TEM grids using focused electron beam induced deposition (FEBID) with a Co$_2$(CO)$_8$ precursor, using a Thermo Fisher HELIOS-600 Dual beam SEM+FIB.
The geometries were defined as 3D \texttt{stl} files, and transformed to stream files, which are readable to the microscope, using the open-access code \texttt{f3ast} \cite{2020Skoric, f3ast_github}. 
Accelerating voltages of \SI{5}{kV} and a beam current of \SI{43}{pA} at a chamber pressure of \SI{3e-6}{mbar} at growth was used, leading to initial growth rates of $\text{GR}_0=\,$40–70 \unit{nm \,s^{-1}}. 
A variable dwell time increasing with growth height was used to account for thermal effects during the nanowire growth \cite{2020Sanz-Hernandez, f3ast_github}.

The individual helix strands have a diameter of \SI{85}{nm}, while the helix is characterized by a periodicity of \SI{250}{nm} and a nominal gyration radius of \SI{34}{nm} (corresponding to a strand overlap of \SI{18}{nm}). These parameters are similar to previous works \cite{2020Sanz-Hernandez,2025Fullerton}. 
Each chiral section is about \SI{750}{nm} long, with the left-handed segment on the bottom standing on a short straight pedestal. 

\subsection{Magnetic Soft X-ray Nanotomography}
\label{sec:methods:alba}

Magnetic vector tomography was performed at the MISTRAL beamline of the ALBA synchroton, with x-rays tuned to the Co $L_2$ absorption edge at a nominal energy of \SI{794.8}{eV}. 

\textbf{Bloch point initialization.}
\label{sec:metods:ALBA:init}
The Bloch-point state was initialized ex-situ by applying a magnetic field along the $x$ direction of the nanostructures of about \SI{1.2}{T} using an electromaget. After the field was reduced to zero, the sample was transferred to the beamline X-ray transmission microscope.

\textbf{Data acquisition.} The lateral resolution of the microscope ($\approx$ \SI{30}{nm}) is largely determined by the objective Fresnel zone plate lens used \cite{2016Oton}. 
A magnification of x1300 was employed, leading to an image pixel size of \SI{10}{nm}.  
%
To reconstruct the 3D magnetization $\vec{m}(\vec{r})$, images from two tilt series were recorded: First, for rotations around the main axis of the nanowire 
from \ang{-65} to \ang{+64} in \ang{1.5} steps, and second, for tilts around the $y$ axis from \ang{-33} to \ang{+19} in \ang{1} steps. 
%
At each angle, measurements were taken with left- and right-handed polarization, $C_L$ and $C_R$, respectively. Flat field images with the sample removed from the imaging path were measured every hour for both polarizations; these allow correction for inhomogeneous illumination and quantitative determination of sample transmittance before calculating the structural and XMCD contrast (see Supplemental Information Sec.~\ref{ssec:methods:alba:tomo} for image processing details). 
For each projection angle and x-ray polarization 40 images (with a total exposure time of \SI{120}{s}) were taken and subsequently averaged to increase the signal-to-noise ratio. 

\textbf{Structural and magnetic contrast.}  The post-processing of the data to compute the xray transmittance and correct for image shifts in order to obtain averaged transmission images with left- and right-handed xray polarization, $T_L$ and $T_R$,
are described in Supporting Information Sec.~\ref{ssec:methods:alba:tomo}. 
The relative shift between features in $T_L$ and $T_R$ are obtained from image registration, with $T_R^\prime$ being the shifted $C_R$ image. From this, the structural contrast (optical density) and XMCD contrast corresponds to the average and difference of the natural logarithm of the two images, respectively:
%
\begin{eqnarray}
	\text{OD} & = & \frac{1}{2} \log{\left( T_L \, T_R^\prime \right)} \, , \\ 
	\text{XMCD} & = & \frac{1}{2} \log{\frac{T_R^\prime}{T_L} } \, .
\end{eqnarray}
%

\textbf{Tomographic Reconstruction.}
\label{sec:methods:alba:reconstruction}
The aligned structural contrast and XMCD tilt series were used as input to perform the  structural and magnetic vector reconstructions using a Simultaneous Iterative Reconstruction Technique (SIRT) evolved from the one described in Ref.~\cite{2018HierroRodriguez}.
Further details of the reconstruction process are explained in Supporting Information Sec.~\ref{ssec:methods:alba:reconstruction}.
The reconstructions of the scalar electronic density and vector magnetization $\vec{m}(\vec{r})$ were obtained with a voxel size of (\SI{10}{nm})$^3$, result of the pixel size defined by the magnification in the microscope. The reconstruction resolution is estimated to be $\approx$ \SI{50}{nm},
similar to previous experiments \cite{2020Hierro-Rodriguez}.
%
%
%
3D data representation and streamlines computation was done with \texttt{paraview}.

\subsection{Transmission Electron Microscopy}
\label{sec:methods:TEM}

Transmission electron microscopy was performed using an FEI Titan$^3$ 80-300 TEM instrument (ThermoFisher Scientific, US) operated in Lorentz-Mode at an acceleration voltage of \SI{300}{kV}. 

\textbf{Off-axis electron holography.} 
To map the remanent magnetic configuration of the helical nanowires after Bloch point nucleation, off-axis electron holography was employed. In this technique, an electron wave passing through a magnetic specimen acquires an Aharonov-Bohm phase shift with respect to a vacuum reference. This phase shift contains contributions from both the electrostatic potential and the magnetic vector potential of the sample. Holograms were recorded using an electrostatic Möllenstedt biprism biased at $+$250~V, which yields a fringe spacing of 1.2~nm. The holographic images were captured with a Gatan OneView CMOS camera, providing a fringe contrast of approximately 10~$\%$ in vacuum. In order to disentangle the electrostatic and magnetic contributions to the measured phase shift, each hologram was acquired twice: first with the specimen in its original orientation and then after flipping the sample upside-down. The sum of the two phase reconstructions yields twice the electrostatic phase, whereas their difference provides twice the magnetic phase, enabling a quantitative separation of the two components. In most cases, the holograms were recorded at an angle of \ang{-40} [see Figs.~\ref{fig:Fig2-TEM-yz}(b,c)], while the nanowire kink points out-of-plane at an angle of \ang{0}.

\textbf{Bloch point initialization}
\label{sec:metods:TEM:init}
To nucleate a Bloch point in the double-helical Co nanowires, an external saturation field of about \SI{0.3}{T} was applied by exciting the objective lens (which is normally switched off in Lorentz mode imaging). 
%
The nanowire was first aligned with the TEM goniometer so that the external field direction lies in the plane of the kink orthogonal to the wire axis (i.e., $H_\text{ini}\parallel x$). This configuration ensures that the magnetization is driven into a saturated state transverse to the wire.
After saturation, the field was gradually reduced to zero. The specimen was then rotated back to its original orientation, and the remanent magnetic configuration was recorded at zero applied field.

\textbf{Electron holographic tomography.} 
Two tilt series of off-axis electron holograms were acquired over a tilt range of \ang{-71} to \ang{+69} in \ang{2} increments, aligning the tilt axis along the nanowire axis. 
After acquiring the first series, the specimen was flipped upside-down outside the microscope, and a second series was recorded under identical imaging conditions, providing the opposite magnetic polarity necessary for separating electric and magnetic phase shifts and hence fields.  
The holograms were reconstructed by Fourier filtering to yield two tilt series of phase images. The latter were subjected to the alignment protocol described by Wolf~et~al.~\cite{2015Wolf}, which included, \textit{e.g.}, correcting for image displacements via cross-correlation and center-of-mass approach within a tilt series, registering the tilt-axis, and correcting for displacements between the original and flipped series. From the added and subtracted phase maps the electrostatic phase ($\propto V(x,y,z)$) and the magnetic phase ($\propto B_z(x,y,z)$) were extracted, giving access to the three-dimensional electrostatic potential $V(x,y,z)$ (structural information) and the magnetic-induction component parallel to the tilt axis, $B_z(x,y,z)$.  
For the 3D reconstruction, the aligned phase stacks were fed into the weighted simultaneous iterative reconstruction technique (W-SIRT) \cite{2014Wolf_a} to obtain quantitative tomograms of both $V(x,y,z)$ and $B_z(x,y,z)$.
More details about the TEM measurements and data analysis are provided in Supporting Information Sec.~\ref{ssec:methods:tem}.

\textbf{Relation between electron and xray reconstruction of the magnetization.}
The magnetic contrast mechanisms used in this work are fundamentally different, with electron holography having non-local sensitivity to the magnetic induction $\vec{B}$, whereas XMCD is proportional to the local magnetization $m\parallel k$. Regardless, both methods ultimately probe the same physical quantity, since XMCD is sensitive only to the solenoidal (divergence-free) part of the magnetization vector field which is equivalent to $\vec{B}$ \cite{2018Donnelly}. Our measurements thus provide direct access to the three-dimensional spin texture forming the spin texture of the Bloch point, with the singular core remaining below the experimental resolution.




\subsection{Micromagnetic Simulations}
\label{sec:methods:sims}
 
Finite-element simulations were run using custom code \texttt{magnum.pi} \cite{2013Abert}.
The double-helix geometry shown in Fig.~\ref{fig:BPs}(e) was defined using python-based \texttt{cadquery}, and then segmented into a triangulated mesh using open-source software \texttt{gmsh} and \texttt{salome}.

Magnetic material parameters were chosen as $M_\text{sat}=\SI{700}{kA\, m^{-1}}$, $A_\text{ex}=\SI{10}{pJ\, m^{-1}}$, $K=0$, similar values to previous works with FEBID cobalt \cite{2021Donnelly}. $\alpha=1$ was used to reduce computation times since we focus here on equilibrium states. 
To obtain the remanent spin state of a Bloch point, the initial magnetic state was set to be saturated along the $x$ direction, and then a magnetic field $H_x$ parallel to the initial magnetization was reduced from $\pm\SI{180}{mT}$ to zero in \SI{0.5}{mT} steps. 
%
The LLG time evolution was performed in discrete time steps of \SI{10}{ps}, with at least ten steps at each field to relax towards a equilibrium spin state.

Subsequent analysis of the simulated magnetic state and its topological properties has been performed using the python-based package \texttt{pyvista}. Derived quantities from the magnetization $\vec{m}(\vec{r})$ include the magnetostatic volume charge, given by the divergence  $\rho_m=\nabla\cdot\vec{m}(\vec{r})$ \cite{2016Arrott}, and the emergent vorticity field $\vec{\Omega}(\vec{r})$ \cite{2021Donnelly}
\begin{equation}
	\Omega_i(\vec{r}) = \frac{1}{8\pi}
		\epsilon_{ijk} \epsilon_{rst} m_r \partial_j m_s \partial_k m_t
	\, ,
	\label{eq:methods:ALBA:emergent_field}
\end{equation}
where $\epsilon_{ijk}$ and $\epsilon_{rst}$ denote the antisymmetric Levi-Civita tensor, and indices $i,j,k,r,s,t=\{x,y,z\}$.

%
%

\section*{Acknowledgments}

We thank Isabel Rivas and Laura Casado (LMA, INM, University of Zaragoza, Spain) for technical support for FEBID growth of the nanostructures, and Elina Zhakina for FIB-cutting the TEM grids.
The XMCD tomography experiments were performed at the MISTRAL beamline of the ALBA Synchtroron (Spain) in collaboration with ALBA staff, and supported by Dr David Raftrey. 
%
%
XMCD ptychography results presented in the supplemental material were performed at the HERMES beamline of the SOLEIL synchtron (France), with support of Dr Stefan Stanescu. 
N.L.\ acknowledges funding from UKRI via the Future Leader Fellowship MR/X033910/1 (LIONESS). 
This work was supported by the European Community under the Horizon 2020 program, Contract No. 101001290 (3DNANOMAG), as well as by the Austrian Science Fund (FWF) [10.55776/PIN1629824].
A.\ H.-R.\ acknowledges the support of Spanish MCIN/AEI/10.13 039/50 110 0011 033/ FEDER, UE under Grant PID2022-13 6784NB and of Agencia SEKUENS (Asturias) under grant UONANO IDE/2024/000 678 with the support of FEDER funds.
D.W., O.Z., and A.L.\ acknowledge financial support by the Collaborative Research Center SFB 1143 (project-id 247310070).
A.E.\ H.-A.\ acknowledges the support from the Severo Ochoa Predoctoral Fellowship Program (Nos. PA-23-BP22-093) from the Government of the Principality of Asturias (Spain).
P.F.\ acknowledges support from the  U.S. Department of Energy, Office of Science, Office of Basic Energy Sciences, Materials Sciences and Engineering Division under Contract No. DE-AC02-05-CH11231 (NEMM program MSMAG).

%
%

\printbibliography

\cleardoublepage

{\centering
    \begin{minipage}[b][1.75in]{3.25in}
    \includegraphics{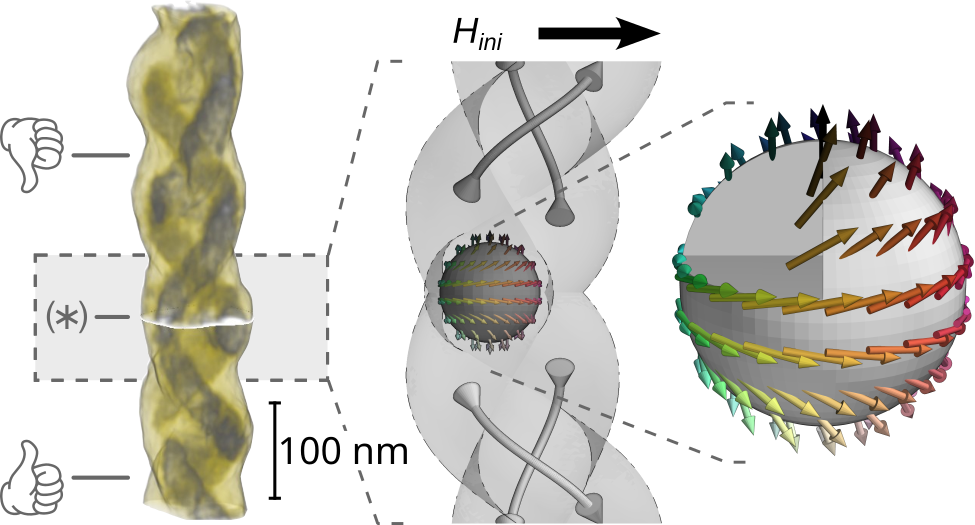}
        
          
    \end{minipage}%

}

\beginsupplement
\title{\papertitle\newline Supporting Information}
\maketitle


\section{XMCD Tomography}

\subsection{X-ray Image Post-Processing}
\label{ssec:methods:alba:tomo}

As described in Methods Sec.~\ref{sec:methods:alba}, for each sample rotation and tilt angle x-ray images $I_{p,j}$, with $p$ denoting the xray polarisation, i.e., $p=C_L$ or $p=C_R$, respectively, 
%
and $j=1\ldots 40$ the image index of equivalent images taken at a given setting to improve the signal-to-noise ratio. 
Additionally, regular flat field exposures $F_{i,j}$ with the sample removed from the light path were taken to correct for an inhomogeneous background illumination and to compute the transmittance.

In the following, we use the symbol $\left<\cdot\right>$ to denote the median, calculated using the function \texttt{numpy.nanmedian}. In particular, $\left<\cdot\right>_j$ denotes the pixel-by-pixel median over a stack of 40~images. We found that this function produced better statistics compared to the normal average (i.e., mean).

\textbf{Calculation of transmittance.} The background illumination in full-field transmission x-ray microscopy is not homogeneous because of the shaking of the capillary condenser; the background also shows slight fluctuations in overall intensity over time. 
To correct for these spatial and temporal background fluctuations, for a given polarization $p$, first we obtain the corrected intensity, respectively transmittance, $T_{p,j}^\text{corr}$ in which the background signal is normalized to $1$ via the following equation:
\begin{equation}
	T_{p,j}^\text{corr} = \frac{I_{p,j}}{f_p \left<F_{p,j}\right>_j} \, .
	\label{seq:im_normalised}
\end{equation}
The normalization factor $f_p$ can be calculated via the ratio of the image and the flat field at the same polarization $p$, where the "mask" selects all pixels that constitute the background, i.e., which lie outside of the structures where the transmittance is unity (here, the mask was implemented using an intensity threshold function):
\begin{equation}
	f_p = \left< \left. \frac{I_{p,j}}{ \left<F_{p,j}\right>_j} \right|_\text{mask}  \right> \, .
\end{equation}

\textbf{Image registration.} Relative drifts between $T_{p,j}^\text{corr}$ images at $j=0$ and $j$ were obtained using a combination of an edge-filter (using \texttt{skimage.feature.canny}) and a Fourier-based cross-correlation algorithm with sub-pixel precision \texttt{skimage.registration.phase\_cross\_correlation} \cite{2008GuizarSicairos}. 
%
%
The average $T_p$ (i.e., $T_L$ and $T_R$ for both polarisations) with the background normalized and images shifted to a common feature $T_{p,j}^{\text{corr}\prime}$ can then be calculated as
\begin{equation}
    T_p = \left<  T_{p,j}^{\text{corr}\prime}  \right>_j
\end{equation}

\textbf{Structural and magnetic xray contrast.} Similar to the previous step, the relative shift between features in $T_L$ and $T_R$ are obtained from image registration, with $T_R^\prime$ being the shifted $C_R$ image.
From this, the optical density (i.e., the electronic or structural contrast) and XMCD contrast corresponds to the average and difference of the natural logarithm of the two images, respectively:
\begin{eqnarray}
	\text{OD} & = & \frac{1}{2} \log{\left( T_L \, T_R^\prime \right)} \, , \\ 
	\text{XMCD} & = & \frac{1}{2} \log{\frac{T_R^\prime}{T_L} } \, .
\end{eqnarray}
%
In this notation, a positive XMCD contrast indicates magnetic moments pointing opposite to the x-ray $k$ vector, i.e., towards the viewer.
%

\begin{figure*}[bt]
	\centering
	\includegraphics[width=0.85\textwidth]{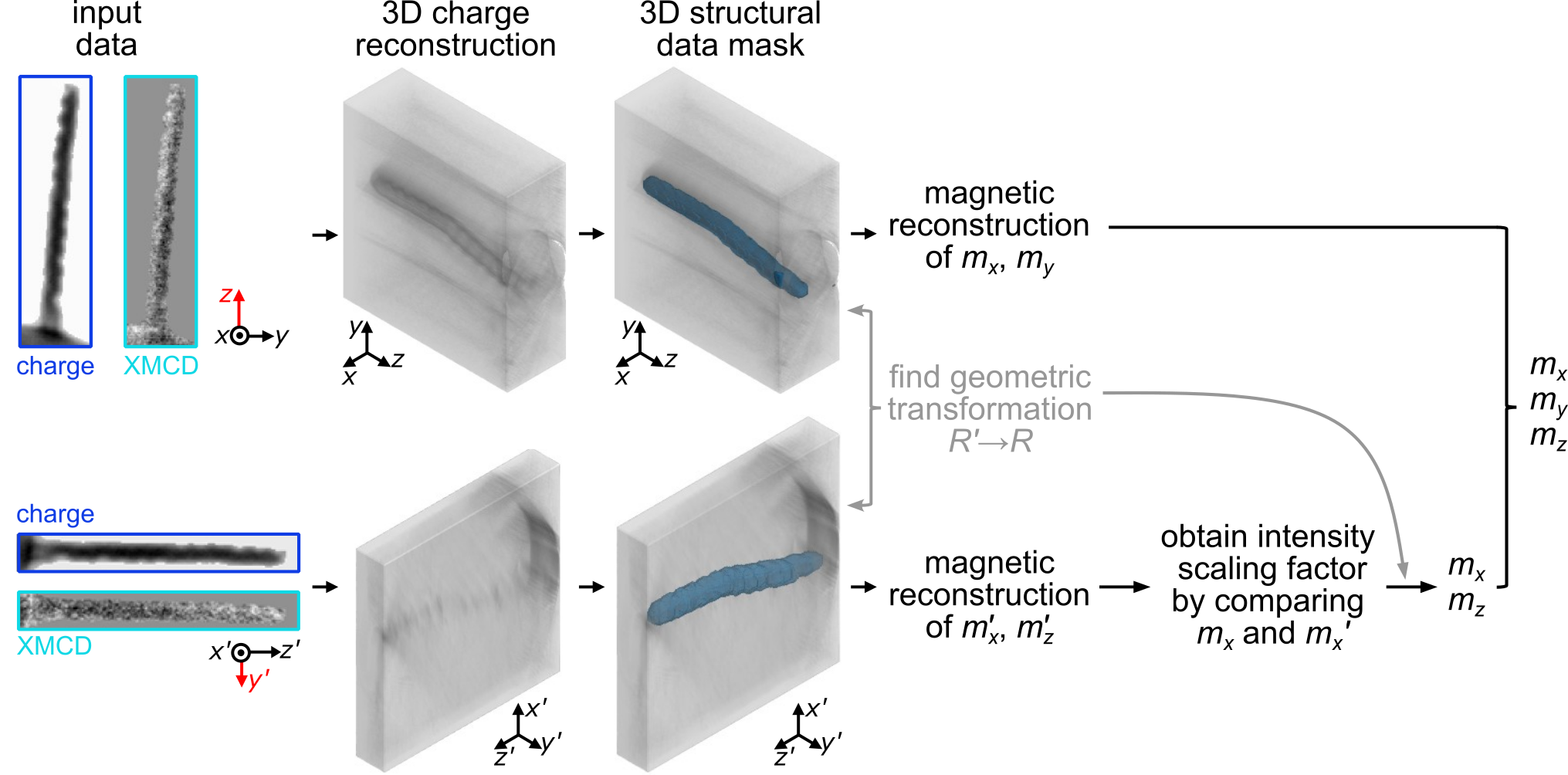}
	\caption{%
		\textbf{XMCD tomography reconstruction workflow.}
        See Sec.~\ref{ssec:methods:alba:reconstruction} for details.
		}
	\label{sfig:ALBA_tomo_workflow}
\end{figure*}

\subsection{Tomographic Reconstruction}
\label{ssec:methods:alba:reconstruction}
\label{ssec:methods:alba}


\textbf{Image series alignment and masking.} All images in the two series, i.e., rotation around the $z$ axis of the nanostructure and tilts around the $y^\prime$ axis (which is roughly parallel to $x$, but not necessarily perfectly perpendicular to $z$), were aligned manually to each other. 
Using the structural (charge) contrast, masks were created for each charge and XMCD image, to limit the input data for the reconstruction.

\textbf{Tomographic reconstruction.} The workflow of the XMCD tomographic reconstruction is illustrated in Fig.~\ref{sfig:ALBA_tomo_workflow}. The aligned structural contrast and XMCD tilt series were used as input to perform the structural and magnetic vector reconstructions using a Simultaneous Iterative Reconstruction Technique (SIRT) evolved from the one described in Ref.~\cite{2018HierroRodriguez}. For both charge and magnetic reconstructions 20 SIRT iterations were employed.

\textbf{Reconstruction of charge density.} First, the scalar density was reconstructed from the charge contrast data allowing to obtain the sample geometry. 
The two experimental rotation axes may neither be fully aligned with the coordinate system of the nanostructure, and, more importantly, not necessarily perfectly perpendicular to each other. Therefore, a geometric transformation is found that maps the reconstructed charge density of each series onto the other.

\textbf{Reconstruction of 3D magnetization.}~The magnetic information is computed from the XMCD images constraining the reconstruction volume to the geometry obtained from the charge density 3D  reconstruction.
%
As XMCD is only sensitive to the magnetization component parallel to the $k$ vector, i.e., $m_k\parallel k$, for each series two components of the solenoidal part of the three-dimensional magnetization can be reconstructed. These are $m_x$ and $m_y$ for the rotation around $z$, and $m_x^\prime$ and $m_z^\prime$ for the rotation around the $y^\prime$ axis.
As the values of $m_x$ and $m_x^\prime$ should be the same, these are compared to find a global conversion factor which allows to put in the same scale $m_z^\prime$ from the second tilt series.
Finally, using the previously determined geometric transformation $R^\prime \rightarrow R$, the magnetization components in the frame of reference of the first tilt series can be computed obtaining $m_x$, $m_y$, and $m_z$. 

\section{Curvature-Driven Bloch Point Displacement}
\label{ssec:curvature_driven_displacement}

\begin{figure}
	\centering
	\includegraphics{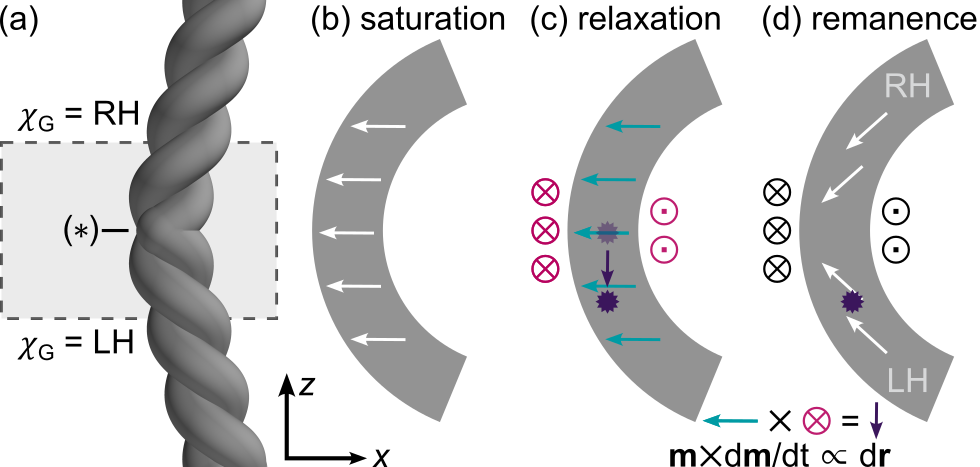}
	\caption{
		\textbf{Deterministic shift of a Bloch point domain wall}, in a curved nanowire, as a result of the damping term of the Landau-Lifschitz-Gilbert equation. 
        (a)~Double-helix geometry with a chirality interface at $(\ast)$. For this explanation, the kink is simplified to a segment of constant curvature.
        (b)~Field-set saturated state with $\vec{m}\parallel x$.
		(c)~During magnetic relaxation, the emergence of circulation at the curved segment leads to an imbalance of $\frac{dm_y}{dt}$ contributions, leading to a net torque $\tau_z \propto m_x\times\frac{dm_y}{dt}$ acting on the Bloch point domain wall.
        (d)~As the combination of the initializing field and the geometric control imposes a deterministic circulation to emerge, the Bloch point displacement is deterministic towards the left-handed (right-handed) segment for a head-to-head (tail-to-tail) configuration.
        }
	\label{sfig:displacement}
\end{figure}


As shown previously \cite{2025Ruiz-Gomez}, Bloch point domain walls are energetically unfavorable in curved nanowire sections. 
The field initialization procedure illustrated in Fig.~\ref{fig:BPs} implies nucleation of a Bloch point at or close to the chirality interface placed at the kink in the helical nanowire. 
Therefore, it can be expected that the Bloch point will displace away from the kinked region upon reducing the magnetic field to zero. 

The two Bloch points selected by the geometry of the helical nanowire shown in Figs.~\ref{fig:BPs}(f,g) -- clockwise head-to-head and counter-clockwise tail-to-tail -- can be transformed into each other by time reversal symmetry, and thus have the same energy at remanence. 
Furthermore, the Bloch points are achiral and have net zero magnetization. This implies that there are no specific symmetry breaking terms to the static energy of the remanent state that explain how the magnetic texture is driven out of the high-curvature region of the nanowire.

\textbf{Arguments for Bloch point displacement via the LLG damping term.} Instead, we find that during the initialization procedure described in Fig.~\ref{fig:BPs}, the coupling between the polarity and its emergent circulation gives rise to an additional torque in a curved wire. As a result the Bloch point is expected to displace from the central chirality interface along a well-defined direction.

In the double-helix geometry shown in \ref{sfig:displacement}(a), the Bloch point domain wall experiences a torque driven by the (second) damping term to the Landau-Lifschitz-Gilbert (LLG) equation:
\begin{equation}
    \frac{\text{d} \vec{m}}{\text{d}t} = -\gamma \vec{m}\times\vec{h}_{\text{eff}} 
    +
    \alpha \vec{m}\times\frac{\text{d}\vec{m}}{\text{d}t}
    \, .
    \label{seq:LLG}
\end{equation}
This additional torque can be intuitively understood using the schematics shown in Fig.~\ref{sfig:displacement}(b-d):
For a high field $\vec{H}\parallel x$, Fig.~\ref{sfig:displacement}(b), an initial saturated state with all moments $\vec{m}\parallel x$ is achieved.
Upon reducing the field to zero, this configuration will relax towards the remanent state shown in Fig.~\ref{sfig:displacement}(d), leading to the stabilization of a Bloch-point domain wall. 
The sign of the initial $H_x$ field implies a head-to-head final configuration is selected in the sketched case. 
Importantly, the geometric design of the helical wires, i.e., the order of left-handed and right-handed wire segments, imposes a specific circulation (here clockwise) that emerges during the relaxation process.
The evolution during the relaxation process, Fig.~\ref{sfig:displacement}(c), is therefore governed by the shape anisotropy, rotating initial $\vec{m}\parallel x$ moments towards the $y$ direction (to establish the circulation) as well as $\pm z$ (to set the polarity). 
As the structure shows a kink (simplified in Fig.~\ref{sfig:displacement} to a segment of constant curvature), and the nanowire diameter is finite, upon reduction of the magnetic field the $m_y$ component shows a transient evolution that is asymmetric, with a curvature-induced net contribution
\begin{equation}
    \left. \frac{|\text{d}m_y|}{\text{d}t} \right|_\text{outer} > \left. \frac{|\text{d}m_y|}{\text{d}t} \right|_\text{inner} \, ,
\end{equation}
leading to a deterministic torque $\tau_z \propto m_x\times\frac{\text{d}m_y}{\text{d}t}$ that displaces via $\text{d}\vec{r}_z\propto\tau_z$ the Bloch point towards the (bottom) left-handed segment for a head-to-head configuration. 

\textbf{Displacement direction.} For a tail-to-tail configuration, both the sign of the initial $m_x$ component reverses as well as the final circulation, leading to an opposite displacement to the right-handed segment. 
Upon reversal of the geometric order of left-handed and right-handed segments, the opposite magnetic circulation is stabilized with applied $H_x$ field, thus also inverting the direction of the displacing torque, resulting in a displacement towards the section of the same geometrical chirality as before.
In conclusion, regardless of the order of the left- and right-handed nanowires, a head-to-head (tail-to-tail) Bloch point will displace towards the left-handed (right-handed) segment.

\textbf{Bloch points in achiral wires.} The relaxation-driven displacement $\Delta z$ of a Bloch point from a high-symmetry point within a curved region conversely allows to \textit{indirectly} indirectly infer the sense of circulation of a Bloch point initialized in \textit{achiral} cuved nanowires \cite{2023Hermosa}. 
Defining the circulation sign as $\sigma(\gamma) = +1$ for counter-clockwise and $-1$ for clockwise circulation when viewed from $+z$ downwards, the displacement obeys $\sigma(\Delta z)\propto\sigma(\gamma)$. Within this LLG-based picture, a deterministic shift of a Bloch point domain wall relative to a curved section is therefore expected.
In experimental systems, however, pinning at local inhomogeneities or small misalignments of the initialization field may dominate over this geometry-induced dynamic displacement. Based on TEM and XMCD imaging data, we argue in the main manuscript that this is indeed the case for our nanostructures, as the observed displacement does not always correspond to the predicted from the damping term of the LLG equation.

\section{Determination of the Helicity Angle \azangle}
\label{ssec:helicity_angle}

\begin{figure}[tb]
	\centering
	\includegraphics[width=172mm]{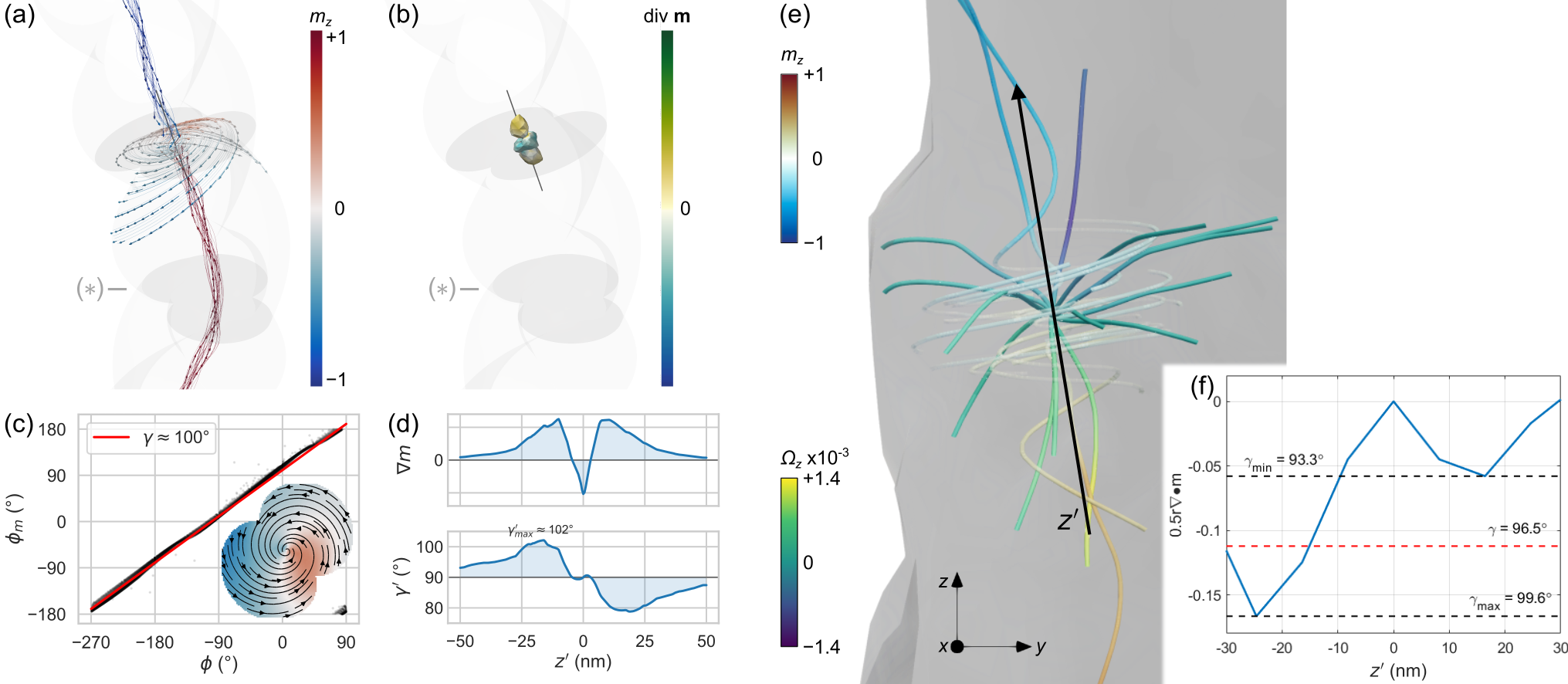}
	\caption{
		\textbf{Determination of helicity angle $\azangle$.}
		(a)~Streamline magnetization plot of a head-to-head Bloch point. Its equatorial plane and main axis are indicated, as well as the chirality interface $(\ast)$.
		(b)~Double-lobed iso-surfaces of the magnetostatic charge $\nabla\cdot\vec{m}$ indicate the Bloch points' hyperbolic character.
		(c)~Plot of $\phi_m$ against $\phi$ within the equatorial plane. The inset shows the spin configuration within the equatorial plane, demonstrating the inwards tilt of the circulating magnetization.
		(d)~Top: Magnetostatic charge $\nabla\cdot\vec{m}$ along the $z^\prime$ axis of the Bloch point. Bottom: Estimation of the helicity angle via Eq.~(\ref{seq:helicity_z}).
        (e,f)~Determination of $\azangle$ from experimental XMCD tomography. (e)~Stream tracer of magnetisation $\vec{m}$ and magnetic vorticity $\vec{\Omega}$ of the experimental configuration. (f)~The variation of Eq.~(\ref{seq:helicity_z}) along the main axis $z^\prime$ of the Bloch point [marked black in (e)], indicates a maximum angle of $\gamma\approx\ang{99.6}$, in agreement with simulations.
		}
	\label{sfig:helicity_angle}
\end{figure}



Figure~\ref{sfig:helicity_angle}(a) shows the spin structure $\vec{m}(\vec{r})$ in the vicinity of the Bloch point, where the two opposing vortex cores meet. The circulating moments indicate the formation of a Skyrmion-tube-like configuration \cite{2025Fullerton} in the segment closer to the chirality interface at $(\ast)$, as the core axial magnetization (in red) opposes the $m_z$ component of the outer spins (in blue).

From the spin state we determined the orientation of the main axis $z^\prime$ of the Bloch point, which is about \ang{3} inclined to the main axis of the helical section tilted at $\theta_\text{tilt}$=\ang{15}. The Bloch point axis as well as the perpendicular equatorial plane are indicated in Figs.~\ref{sfig:helicity_angle}(a,b).
Figure~\ref{sfig:helicity_angle}(b) shows the iso-surfaces of the magnetostatic charge distribution $\nabla\cdot\vec{m}(\vec{r})$, showing a double-lobed configuration around the Bloch point, indicating its hyperbolic character \cite{2024Hermosa-Munoz}.

To determine the value of the helicity angle \azangle\ we used two different approaches.

First, we directly consider the spin configuration in the equatorial plane of the Bloch point, which is shown as inset in Fig.~\ref{sfig:helicity_angle}(c). Using Eq.~(\ref{eq:phi_m}), one can extract $\gamma$ from relating the spin angle $\phi_m$ to the real-space polar angle $\phi$. To determined $\phi$ from the $(x,y)$ coordinates, the  center of mass of the simulated exchange energy $\propto\Sigma_{i,j} (\partial_j m_i)^2$ was used, as this point corresponds the central position of the Bloch point. 
The resulting $\phi_m(\phi)$ graph is shown in Fig.~\ref{sfig:helicity_angle}(c), and from linear regression we obtained value of $\gamma_{\text{sim}}\approx\ang{100}$. The fact that $\azangle>\ang{90}$ indicates an inwards tilt to the circulation, as is evident in the inset where the in-plane moments slightly spiral inwards. 

Second, the helicity angle \azangle\ can be obtained by an approximate formula proposed in \cite{2024Hermosa-Munoz}, involving the magnetostatic charge, respective divergence $\nabla\cdot\vec{m}$, along the main axis $z^\prime$:
\begin{equation}
	\cos(\gamma^\prime) = \frac{z^\prime}{2} \nabla\cdot\vec{m} 
	\label{seq:helicity_z}
\end{equation}
Fig.~\ref{sfig:helicity_angle}(d) shows the evolution of $\nabla\cdot\vec{m}$ along $z^\prime$ on the top, and the term from Eq.~(\ref{seq:helicity_z}) on the bottom. A maximum value of $\gamma_{\text{sim}}^\prime\approx\ang{102}$ is observed very close ($\approx\SI{20}{nm}$) to the Bloch point, matching the value of $\gamma_{\text{sim}}\approx\ang{100}$ from $\phi_m(\phi)$ above. 

We used the second method to perform an equivalent analysis for the reconstructed XMCD tomography data $\vec{m}(\vec{r})$, as shown in Figs.~\ref{sfig:helicity_angle}(e,f); which returned an experimental value of \mbox{$\gamma^\prime_{\text{exp}}\approx\ang{99.6}$}, in agreement with simulations. 
%

\section{2D and 3D Transmission Electron Microscopy}
\label{ssec:methods:tem}

\begin{figure}[b]
	\centering
	\includegraphics[width=0.5\textwidth]{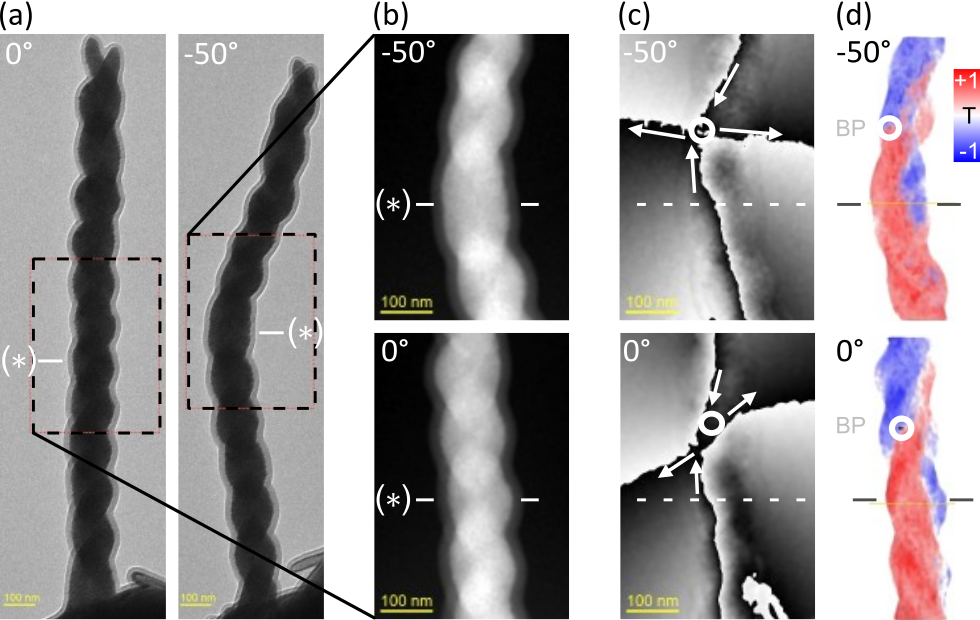}
	\caption{
		\textbf{Electron holographic tomography.}
		(a)~Bright-field transmission electron microscopy images of the double-helix cobalt nanowire. The position of the chirality interface is marked with $(\ast)$, and the chosen field of view for tomographic reconstruction around the kink highlighted with a dashed box. 
        %
        (b)~Electrostatic and (c)~magnetic phase images for different tilt angles.
        In (c), the phase wraps modulo two $\pi$, amplifying equi-phase lines, illustrate the projected $B$-field lines (white arrows), which are consistent with the internal $B_z$ component visualized as red-blue volume rendering in (d). The inferred Bloch point position is indicated by white circles.
		}
	\label{sfig:TEM_Tomo_regions}
\end{figure}

\begin{figure*}[bt]
	\centering
	\includegraphics{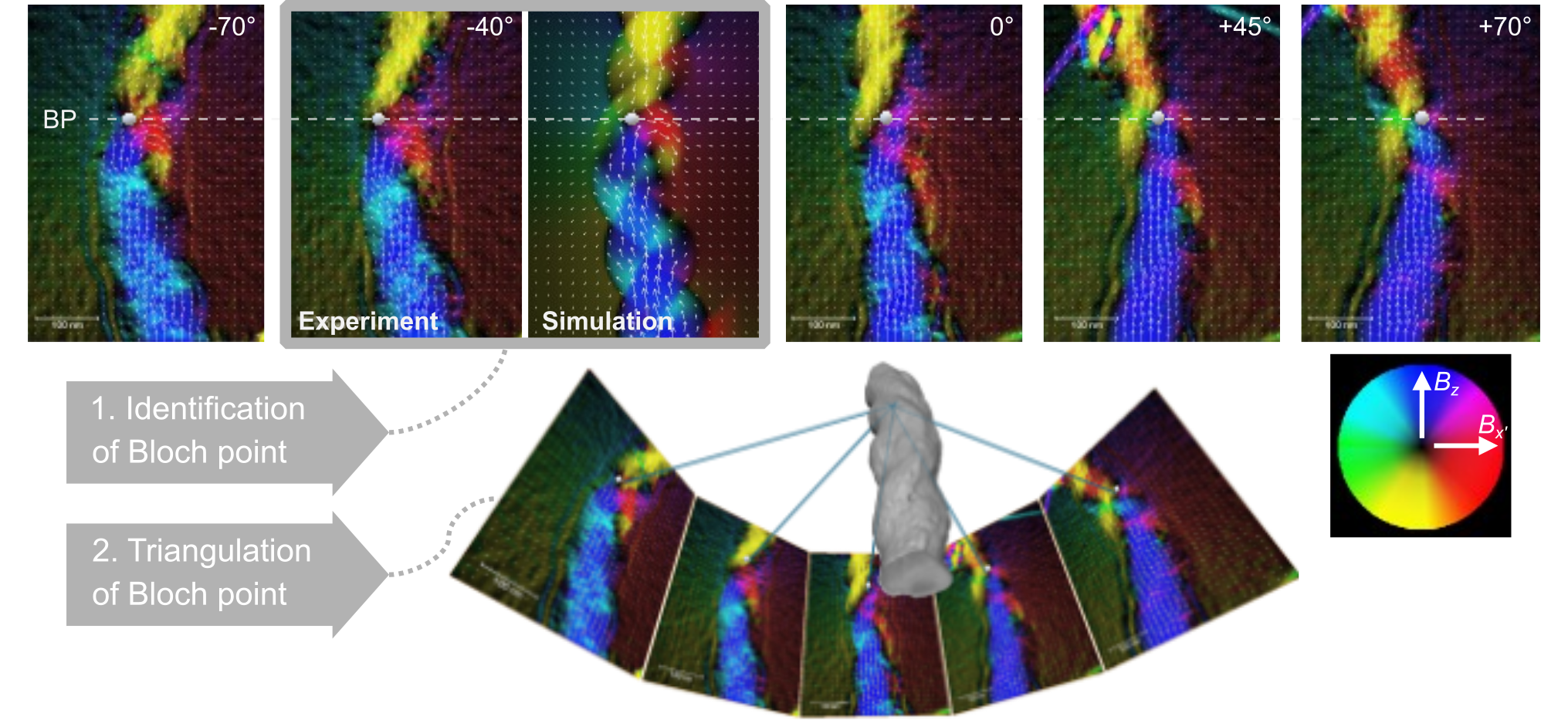}
	\caption{%
		\textbf{Determination of the Bloch point position} from TEM holography. 
        First, comparison with simulation results allows to identify the position of the Bloch point within a specific 2D induction phase map projection. 
        Second, combining the positions from three or more projections allows to triangulate the Bloch point position in 3D.
        }
	\label{sfig:BP-triang}
\end{figure*}


\textbf{Electron holography}. Electron holograms at the position shown in Fig.~\ref{sfig:TEM_Tomo_regions}(a) were recorded and reconstructed to obtain phase images covering \SI{750}{nm} of the entire nanowire (see Fig.~\ref{sfig:TEM_Tomo_regions}(a)) around its chirality interface.
 The phase shift $\varphi$ in the object plane $\left(x,y\right)$ is given by 
	\begin{equation}
		\varphi\left(x,y\right)=\intop_{-\infty}^{+\infty}\left(\frac{e}{\hbar v}V\left(x,y,z\right)-\frac{e}{\hbar}A_{z}\left(x,y,z\right)\right)\mathbf{\mathrm{d}}z.
		\label{eq:TEM-phase-shift}
	\end{equation}
Here, $v$ represents the electron velocity, $V\left(x,y,z\right)$ the 3D electrostatic potential, $e$ the elementary charge, $\hbar$ the reduced Planck constant, and $A_{z}\left(x,y,z\right)$ the 3D $z$-component of the magnetic vector potential. 
We can refer to the first term as electric phase shift $\varphi_{\text{el}}$ and the second term as magnetic phase shift $\varphi_{\text{mag}}$. 

By converting the second integral to the magnetic flux enclosed between interfering paths of object and reference wave, the directional spatial derivatives read
%
	\begin{equation}
		\begin{pmatrix}
			\partial_x\varphi_{\text{mag}}\\
			\partial_y\varphi_{\text{mag}} 
		\end{pmatrix}
		=  \frac{e}{\hbar}\int_{-\infty}^{\infty}
		\begin{pmatrix}
			\hspace{10pt} B_{y}\left(\mathbf{r}\right)\\
			-B_{x}\left(\mathbf{r}\right)
		\end{pmatrix}\mathrm{d}z
		\, ;
		\label{eq:PhaseGrad_vs_B}
	\end{equation}
hence, they are proportional to the projected in-plane components of the magnetic induction $\mathbf{B}$. 

To quantify the electrostatic mean inner potential and the magnetic induction, the electric $\varphi_{\text{el}}$ and magnetic phase $\varphi_{\text{mag}}$ shifts must to be separated. 
This is accomplished by first reversing the magnetic phase shift (e.g., by flipping the sample upside down), and second, reconstructing the phase image of the same object again \cite{1986Tonomura}. Then, the sum (difference) of phase images before and after flipping yields twice the electrostatic (magnetic) phase shift $\varphi_{\text{el}}$ ($\varphi_{\text{mag}}$).
Electric and magnetic phase images within the field of view for \ang{0} and \ang{-50} are shown in Fig.~\ref{sfig:TEM_Tomo_regions}(b,c). As indicated by the white arrows, the magnetic field lines seem to converge to the Bloch point position (white circle).

\textbf{Electron holographic tomography.} To reconstruct the three-dimensional data presented in Fig.~\ref{fig:Fig2-tem-tomo} after the Bloch point initialization, a field of view around the central part of the double-helical Co nanowire was chosen for a tomographic measurement series. This region, as outlined in the bright-field electron transmission microscopy images in Fig.~\ref{sfig:TEM_Tomo_regions}(a) contains a domain wall, as detected by means of the stray field modulation visible in the reconstructed phase images. 
%
%
The tilt series acquisition was performed semi-automatically with an in-house developed software to collect two holographic tilt series consisting of object and object-free empty holograms, the second tilt series after the sample was flipped upside-down outside the microscope. 
For the data shown in Fig.~\ref{fig:Fig2-tem-tomo} the range of both tilt series (at original and flipped sample) was from \ang{-71} to \ang{+69} in \ang{2} steps, around the vertical $z$ axis.
%

To obtain the full phase shift $>2\pi$, the phase images were unwrapped -- automatically by the Flynn algorithm \cite{1998Ghiglia}, or manually at regions where the phase signal was too noisy or undersampled by using prior knowledge of the phase shift (e.g., from adjacent projections) \cite{2014Lubk}. 
Prior to the separation of electric and magnetic phase shifts, the two phase tilt series were processed similarly as reported in Ref.~\cite{2019Wolf}. 

The resulting tilt series of electrostatic phase images divided by ${e}/{(\hbar v)}$ [projected potential $V_{\text{proj}}$, i.e., the first term in Eq.~(\ref{eq:TEM-phase-shift})] was used as input to compute the 3D tomogram of the electrostatic potential by the weighted simultaneous reconstruction technique (WSIRT) algorithm \cite{2014Wolf_a}. 
To reconstruct the internal magnetic state, we removed the stray field outside the nanowire by setting $B_z$ values to zero where the 3D potential $V$ falls below a threshold value of \SI{15}{V} representing the nanowire surface [Fig.~\ref{fig:structure}(c)].  

Following Eq.~(\ref{eq:PhaseGrad_vs_B}), the derivatives of the magnetic phase images in the direction \textit{perpendicular} to the tilt axis to obtain the projections of the the axial $B_z$ component times $e/\hbar$. 
%
Because the computation of the derivatives enhances noise, the magnetic phase images were smoothed slightly by a low-pass filter (e.g. spatial Gaussian filter). 
Finally, the WSIRT algorithm was employed to reconstruct the 3D $B_z$ component from the corresponding tilt series.

For visualization and analysis, the tomograms were denoised by means of a non-linear anisotropic diffusion filter within the Avizo software package (ThermoFisher Scientific Company). Fig.~\ref{sfig:TEM_Tomo_regions} depicts the tomographic reconstruction of $B_z(\vec{r})$ viewed under two different tilt angles.

\textbf{Supporting Movies.} \label{sfig:TEM_movie_chirality_interface}
Visualizing the full 3D reconstructions corresponding to the data shown in Figs.~\ref{fig:structure} and~\ref{fig:Fig2-tem-tomo}; in all cases the Bloch point position is indicated with a black voxel:
\vspace{-0.75em}
\begin{itemize}\setlength\itemsep{-0.25em}
    \item \texttt{3D-pot.mp4}: Volume rendering of the electrostatic potential, corresponding to Fig.~\ref{fig:structure}(d).
    \item \texttt{3D-pot-slicing.mp4} Alternative view on the electrostatic potential of the nanostructure. Left: Showing the surface, corresponding to Fig.~\ref{fig:structure}(c). Right: Cross-sectional slices around the chirality interface, clearly showing the reversal of the helix rotation.
    \item \texttt{3D-Bz.mp4} 3D view of magnetic induction within the nanostructure, corresponding to Fig.~\ref{fig:Fig2-tem-tomo}(c).    
\end{itemize}

\begin{figure*}[bt]
	\centering
	\includegraphics[width=0.8\textwidth]{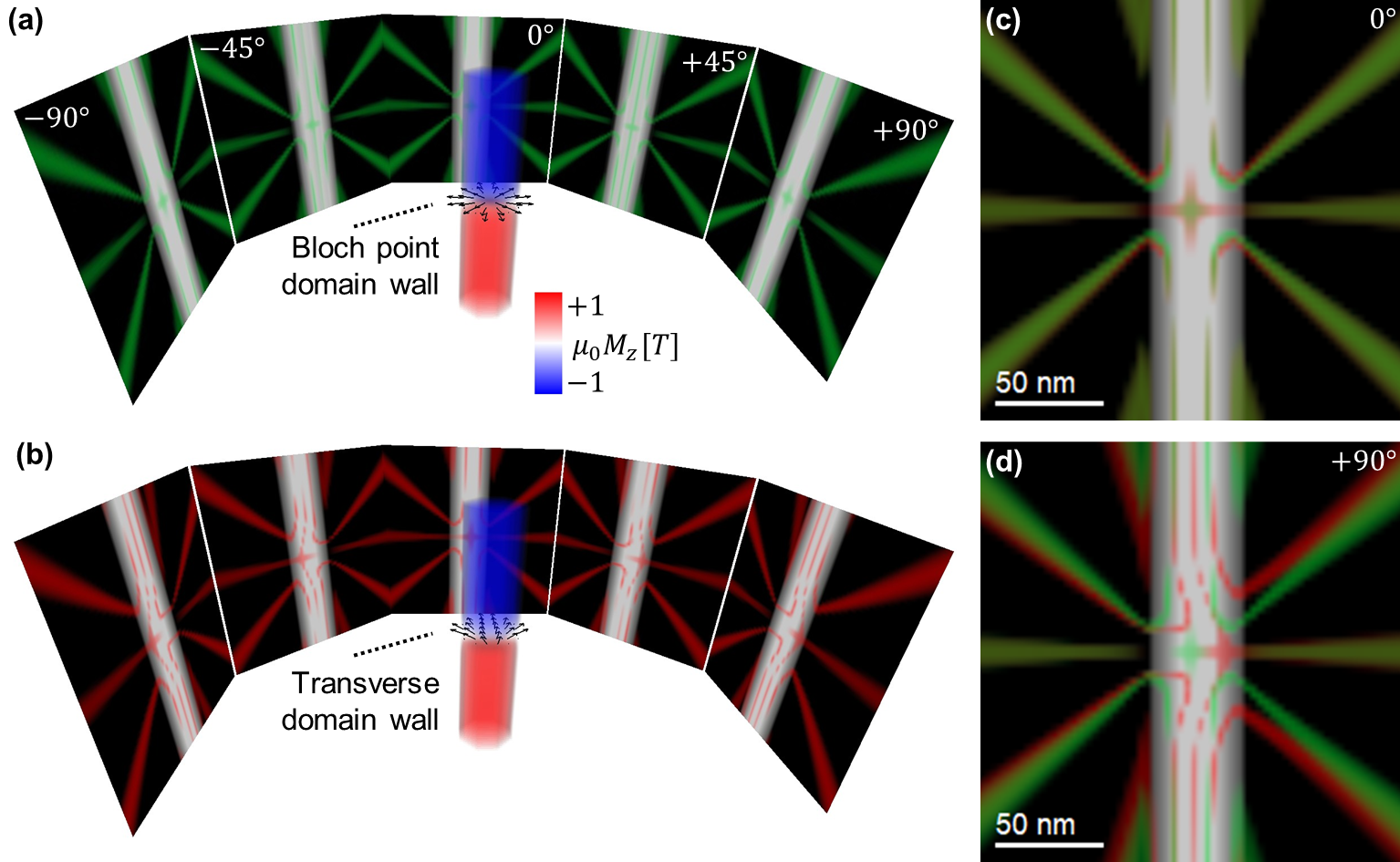}
	\caption{
        \textbf{Stray-fields of Bloch point and transverse domain walls.}
          (a,b) Simulated phase images of a Co nanowire (42~nm diameter) are shown for various projection angles, depicting a Bloch‑point domain wall (a) and a transverse domain wall (b). Electric contributions are rendered in grayscale, while magnetic contributions appear as green iso‑lines (as in Fig.~\ref{fig:Fig2-tem-tomo} of the main text). The stray‑field pattern of the Bloch‑point wall displays monopole‑like symmetry for all viewing directions. In contrast, the transverse wall yields an asymmetric phase image between both sides of the nanowire, with the asymmetry increasing as the angle between the wall’s magnetization and the electron beam grows. Phase images extracted from the tilt series at $0^\circ$ (c) and $+90^\circ$ (d). (c) demonstrates that in both cases Bloch point domain wall (green) and transverse wall (red) the stray field is symmetric, whereas in (d) a pronounced antisymmetry of the stray field for the transverse wall is visible.
        }
	\label{sfig:transverse-wall-TEM-contrast}
\end{figure*}

\textbf{Strayfields of Blochpoint and transverse domain walls obtained in phase image tilt series of simulated nanowires.}
Figure~\ref{sfig:transverse-wall-TEM-contrast} shows the comparison of simulated magnetic phase maps, comparing the observed signatures of Bloch point domain walls and transverse domain walls in magnetic (in our case Co) nanowires. Transverse domain walls are characterised by an asymmetry across the nanowire, which changes with viewing direction. In contrast, Bloch point domain walls exhibit a symmetric magnetic phase contrast that does not depend on the projection direction. This supports the main text’s claim that, based on the observed symmetry of the stray fields in the phase images, a Bloch point exists (see Fig.~2a,b of the main text).

\textbf{Triangulation of the Bloch Point position}
As shown in Fig.~\ref{sfig:BP-triang}, the Bloch-point location can be inferred from a specific 2D projection of the magnetic induction. This is accomplished by comparing the projection with magnetic-induction maps calculated from remanent Bloch-point states obtained via micromagnetic simulations. In the experimental data, the Bloch-point appears as features that match the signatures of the projected Bloch point seen in the simulated maps.
Combining the determined coordinates from three or more different viewing angles allows triangulation of the Bloch point position in three dimensions.

\section{Alternative Bloch Point Initialization Protocol}
\label{ssec:alternative_init_protocol_SOLEIL}

\begin{figure}[tb]
	\centering
	\includegraphics[]{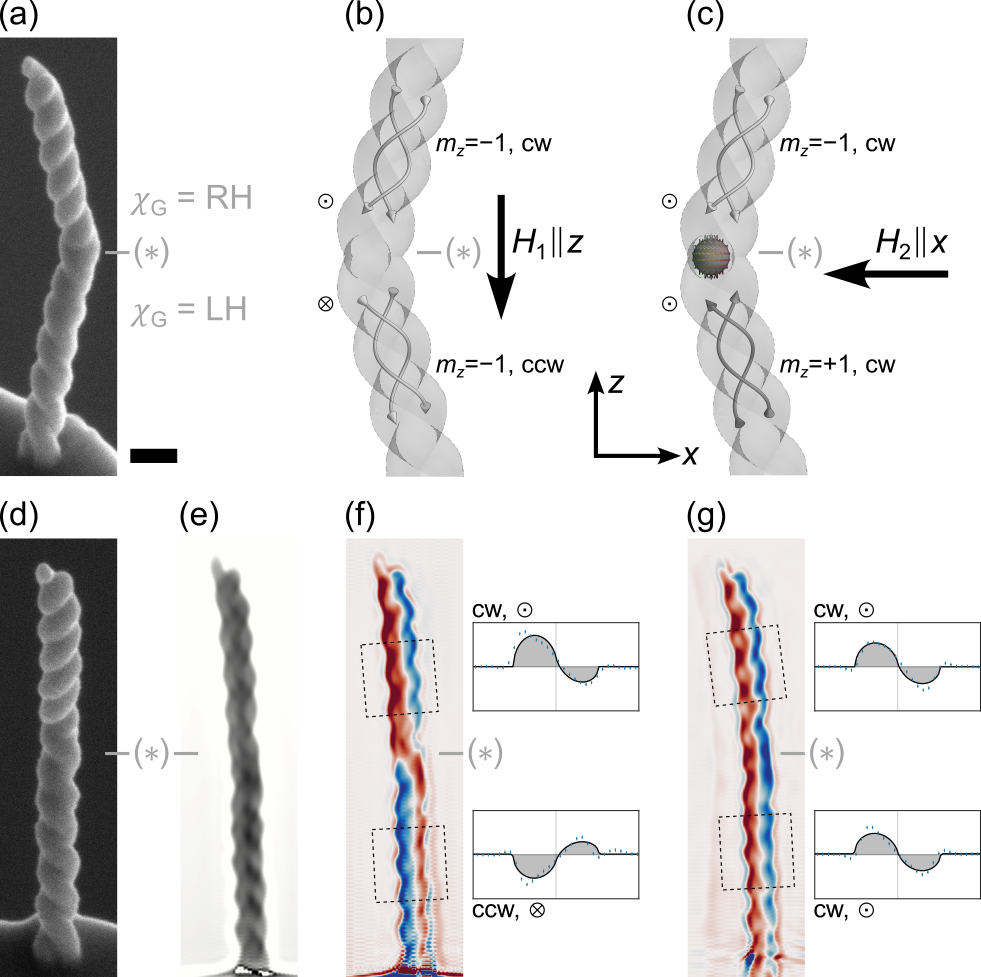}
	\caption{%
			\textbf{In-situ two-step initialization of Bloch points.}
			(a)~SEM side view of structure, with the chirality interface marked with $(\ast)$. The images were taken under a stage tilt of \ang{52}, and the horizontal scale bar measures \SI{150}{nm}. 
			(b,c)~Schematic illustration of the two-step Bloch point initialization protocol, requiring first to set a well-defined axially-magnetized helical state, followed by simultaneously switching the axial magnetization and circulation in one of the segments by a transverse field. 
			(d,e)~Front view of the double-helix structure, measured with (d)~SEM and (e)~x-ray ptychography.
			(f)~Ptychography XMCD image close to remanence after saturation with $H_1=\SI{+180}{mT}\parallel z$. Boxes indicate regions of interest for which average line profiles (to the right) are extracted. Fits to an analytical model allow to extract the sense of circulation and the axial magnetization into or out of the plane ($\otimes$ and $\odot$, respectively).
			(g)~XMCD image at remanence after subsequent application of $H_2=\SI{180}{mT}\parallel x$, showing a uniform circulation across the entire structure and opposite axial magnetization in each section, indicating the formation of a Bloch point close to the chirality interface. 
			%
		}
	\label{sfig:SOLEIL-two-step-ini}
\end{figure}

%

Initializing the Bloch point state with the method illustrated in Fig.~\ref{fig:BPs}(f,g) requires sizable fields to be applied; in our case we used at least \SI{0.3}{T} for the electron holography experiments (Methods Sec.~\ref{sec:metods:TEM:init}) and \SI{1.2}{T} for the x-ray tomography experiments (Methods Sec.~\ref{sec:methods:alba}, using lower fields did not lead to a reliable initialization).
The magnitude of the initialization fields requires them to be applied ex-situ, which is undesirable for future experiments, e.g., probing the field- or current-driven motion of Bloch points, where the initial Bloch point state needs to be re-set regularly.

In Fig.~\ref{sfig:SOLEIL-two-step-ini} we illustrate an alternative two-step protocol to initialize Bloch point domain walls in the helical nanostructure shown in Fig.~\ref{sfig:SOLEIL-two-step-ini}(a). 
First, after application of an $H_z$ field at remanence an axially magnetized state is set. As shown in Fig.~\ref{sfig:SOLEIL-two-step-ini}(b) the circulation in each segment is opposite, governed by the local condition that $\chi_M=\chi_G$. 
Then, a transverse field $H_x$ switches the axial magnetization (and subsequently, the circulation) in one of the two legs only, as shown in Fig.~\ref{sfig:SOLEIL-two-step-ini}(c). After relaxation to zero field, the formation of a Bloch-point domain wall close to the chirality interface is expected.

We demonstrate this field protocol using x-ray ptychography and XMCD experiments at the HERMES beamline of the SOLEIL synchrotron. Details of the general experimental approach, image reconstruction and processing can be found elsewhere \cite{2025Fullerton}. 
%
Most importantly, a sample environment featuring a tuneable array of permanent magnets \cite{2012Nolle} allows to apply magnetic fields up to \SI{180}{mT} along both the beam direction $k\parallel x$, i.e., along the kink of the nanostructure, as well as transverse to $k$, i.e., it along the main $z$ nanowire axis.

Figure~\ref{sfig:SOLEIL-two-step-ini}(d,e) shows the front view of the sample both with SEM as well as the structural x-ray contrast of the nanowire, with the chirality interface marked at $(\ast)$.

Figure~\ref{sfig:SOLEIL-two-step-ini}(f) shows an XMCD image at $H_z=\SI{-10}{mT}$, after application of a field of $H_z=\SI{+180}{mT}$, i.e., corresponding to the first step of the field protocol.
The transverse red-blue XMCD contrast indicates an opposing helical circulation of magnetization in either half of the structure \cite{2025Fullerton}.

We furthermore can fit averaged line cuts of the XMCD contrast (dashed boxes indicate the relevant regions of interest) with an analytical model (black line and shaded area) to extract the sign of the projected axial magnetization $m_z$ as well (see Supporting Information of Ref.~\cite{2025Fullerton} for further details).
For Fig.~\ref{sfig:SOLEIL-two-step-ini}(f) this projected component points into ($\otimes$) and out of ($\odot$) the plane on the bottom and top helix, respectively, indicating an overall continuous axial magnetization.

Figure~\ref{sfig:SOLEIL-two-step-ini}(g) shows a subsequent XMCD image at remanence after application of $H_x=\SI{+180}{mT}$, i.e., after the second step of the field protocol. Here, a uniform circulation across the entire structure is observed and from the fits we conclude that the axial components of the bottom and top segment both point towards the viewer ($\odot$). 
While this observation does not directly probe the local spin structure, it nevertheless indicates the presence of a Bloch-point domain wall, which is expected to form in the vicinity of the chirality interface $(\ast)$.



\printbibliography

\end{document}